%% file: main.tex
\documentclass[acmtog,nonacm]{acmart}
\usepackage{booktabs,amsmath} % For formal tables
\usepackage{xcolor}
\usepackage{graphicx}
\usepackage{tabularx}
\usepackage{makecell}
\usepackage{bm}
\usepackage[normalem]{ulem}
\usepackage{xcolor}
\usepackage{multirow}

\usepackage{xr}
\makeatletter
\newcommand*{\addFileDependency}[1]{
  \typeout{(#1)}
  \@addtofilelist{#1}
  \IfFileExists{#1}{}{\typeout{No file #1.}}
}
\makeatother

\newcommand*{\myexternaldocument}[1]{
    \externaldocument{#1}
    \addFileDependency{#1.tex}
    \addFileDependency{#1.aux}
}
\myexternaldocument{supp}

\input{macros}

\usepackage[ruled]{algorithm2e} % For algorithms

\SetAlFnt{\small}
\SetAlCapFnt{\small}
\SetAlCapNameFnt{\small}
\SetAlCapHSkip{0pt}

\acmJournal{TOG}
\setcopyright{none}
\begin{document}

\begin{teaserfigure}
	\centering
	\setlength{\tabcolsep}{0pt}
	\includegraphics[width=1.0\textwidth]{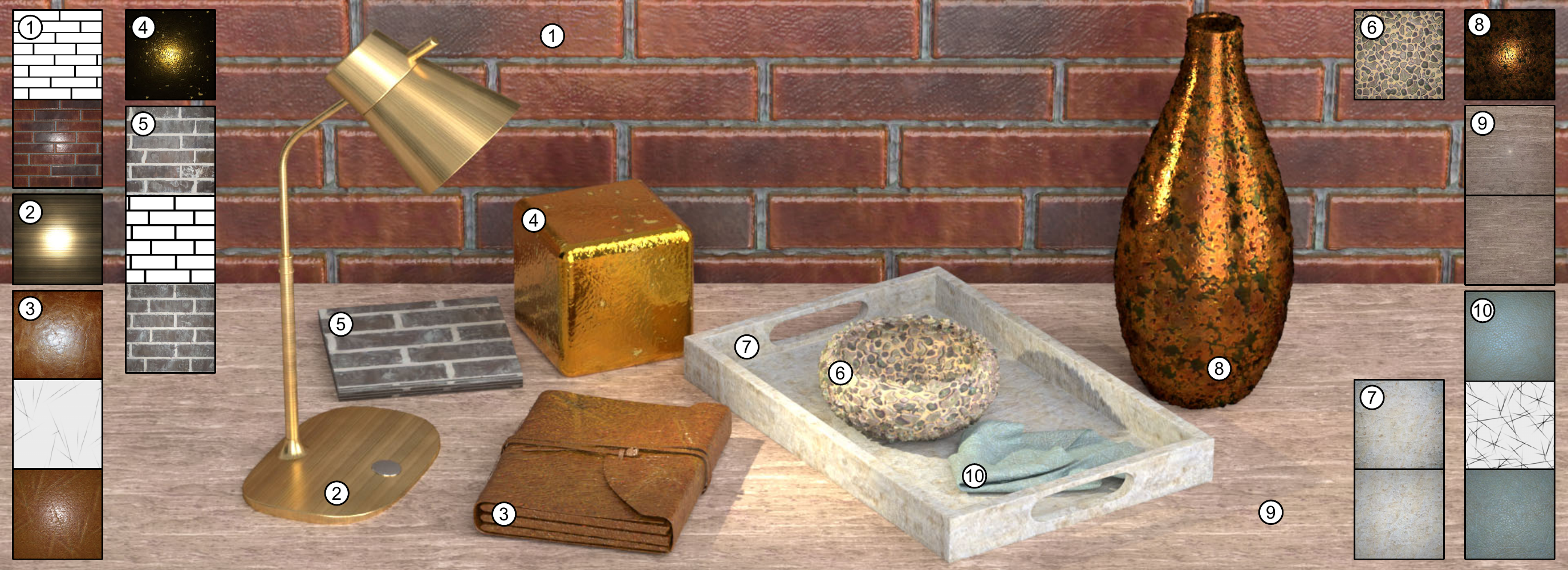}
	\caption{We present a category-specific generative model for spatially-varying materials, whose results are seamlessly tileable and optionally conditioned on patterns controlling the material structure. Our model can be used to generate new materials, or inverted to find materials matching target photographs by optimization. Here we show examples of tile, leather, stone and metal classes, either directly generated (1, 2, 4, 6, 8) or reconstructed from photographs (3, 5, 7, 9, 10) and, in some cases, conditioned on an input structure pattern (1, 3, 5, 10). The insets show, where applicable, the target photograph, condition pattern and a rendering of the synthesized material. The generated maps include height fields that are displacement-mapped in the rendered scene.}
	\label{fig:teaser}
\end{teaserfigure}

% Title portion
\title{\cmgan: Tileable, Controllable Material Generation and Capture}

% DO NOT ENTER AUTHOR INFORMATION FOR ANONYMOUS TECHNICAL PAPER SUBMISSIONS TO SIGGRAPH 2019!

\author{Xilong Zhou}
\affiliation{%
 \institution{Texas A\&M University}
%   \city{College Station} 
%  \state{TX} 
 \country{USA}
}
% \email{zhouxilong199213@tamu.edu}

\author{Milo\v s Ha\v san}
\affiliation{%
 \institution{Adobe Research}
%   \city{San Jose} 
%  \state{CA} 
 \country{USA}
}
% \email{mihasan@adobe.com}

\author{Valentin Deschaintre}
\affiliation{%
 \institution{Adobe Research}
%   \city{London} 
%  \state{CA} 
 \country{UK}
}

\author{Paul Guerrero}
\affiliation{%
 \institution{Adobe Research}
%   \city{London} 
%  \state{CA} 
 \country{UK}
}

\author{Kalyan Sunkavalli}
\affiliation{%
 \institution{Adobe Research}
%   \city{San Jose} 
%  \state{CA} 
 \country{USA}
}

\author{Nima Kalantari}
\affiliation{%
 \institution{Texas A\&M University}
%   \city{San Jose} 
%  \state{CA} 
 \country{USA}
}

\begin{abstract}
\input{abstract}

\end{abstract}

% <concept_id>10010147.10010371.10010372</concept_id>
% <concept_desc>Computing methodologies~Rendering</concept_desc>

% <concept_id>10010147.10010371.10010372.10010374</concept_id>
% <concept_desc>Computing methodologies~Ray tracing</concept_desc>
% <concept_id>10010147.10010257.10010293.10010294</concept_id>
% <concept_desc>Computing methodologies~Neural networks</concept_desc>
% <concept_id>10010147.10010371.10010382.10010383</concept_id>
% <concept_desc>Computing methodologies~Image processing</concept_desc>

% \ccsdesc[500]{Computing methodologies~Rendering}
% \ccsdesc[500]{Computing methodologies~Ray tracing}
% \ccsdesc[500]{Computing methodologies~Neural networks}
% \ccsdesc[500]{Computing methodologies~Image processing}


%\keywords{materials, GAN, conditional}

\maketitle

\input{intro}

\input{related}

\input{forward}
\input{inverse}
\input{results}

\input{conclusion}

\section*{Acknowledgements}

We thank Krishna Kumar Singh, Yijun Li and Jingwan Lu for help with CollageGAN set up and training details.

\bibliographystyle{ACM-Reference-Format}
\bibliography{references}

\input{supple}

\end{document}

%% file: macros.tex
\newcommand{\cmgan}{\textsc{TileGen}\xspace}

\newcommand{\latent}{\bm{u}}
\newcommand{\latentopt}{\bm{u^{*}}}

\newcommand{\img}{\bm{I}}
\newcommand{\albedo}{\bm{a}}

\newcommand{\height}{\bm{h}}
\newcommand{\rough}{\bm{r}}
\newcommand{\pat}{\bm{p}}

\newcommand{\loss}{\mathcal{L}}

\newcommand{\render}{\mathcal{R}}

\newcommand{\generator}{\mathcal{G}}

\DeclareMathOperator*{\argmin}{arg\,min}

\newcommand{\bw}{\bm{w}}
\newcommand{\bz}{\bm{z}}
\newcommand{\noise}{\bm{\xi}}
\newcommand{\calW}{\mathcal{W}}

\newcommand{\calN}{\mathcal{N}}

%% file: abstract.tex
%Authoring photorealistic materials (SVBRDFs) is a time-consuming process that requires specialized skills, such as construction and editing of complex procedural node graphs. 
%Generative adversarial networks (GANs) could be an efficient alternative to generate complex materials; they are also naturally differentiable, which is key for inverse estimation of material parameters from real examples. However, previous approaches for generating materials using GANs are unconditional: they do not have any mechanism for controlling and editing the generated materials. They also suffer from the lack of tileability.
%
Recent methods (e.g. MaterialGAN) have used unconditional GANs to generate per-pixel material maps, or as a prior to reconstruct materials from input photographs.
These models can generate varied random material appearance, but do not have any mechanism to constrain the generated material to a specific category or to control the coarse structure of the generated material, such as the exact brick layout on a brick wall. Furthermore, materials reconstructed from a single input photo commonly have artifacts and are generally not tileable, which limits their use in practical content creation pipelines.
We propose \cmgan, a generative model for SVBRDFs that is specific to a material category, always tileable, and optionally \emph{conditional} on a provided input structure pattern. 
\cmgan is a variant of StyleGAN whose architecture is modified to always produce tileable (periodic) material maps. In addition to the standard ``style'' latent code, \cmgan can optionally take a condition image, giving a user direct control over the dominant spatial (and optionally color) features of the material. For example, in brick materials, the user can specify a brick layout and the brick color, or in leather materials, the locations of wrinkles and folds. Our inverse rendering approach can find a material perceptually matching a single target photograph by optimization. This reconstruction can also be conditional on a user-provided pattern. The resulting materials are tileable, can be larger than the target image, and are editable by varying the condition.

%% file: intro.tex
\section{Introduction}

High-quality materials are critical to realism in computer graphics applications. While reflectance models and rendering algorithms have reached excellent fidelity over the last decade, authoring photorealistic materials is still a time-consuming process that requires the construction and editing of complex procedural node graphs, and/or skilled manipulation of textures captured from photographs.

Given the rapid progress in generative adversarial networks (GANs) \cite{StyleGAN, StyleGAN2, StyleGAN3}, such models are a natural approach for material generation. As an example, MaterialGAN \cite{Guo2020} is a recently proposed generative SVBRDF model trained on a large example dataset of synthetic material maps. MaterialGAN learns a latent space that can be sampled to produce high-quality materials, or leveraged as an optimization prior for SVBRDF capture from photographs. However, MaterialGAN is not tileable and does not have any mechanism for controlling the generated materials. If a user is interested in changing the layout of a synthesized brick texture, there is no simple way to do so other than by repeatedly generating a random material until the desired layout is found by chance. Additionally, the results may have artifacts such as specular highlights leaking into the estimated maps.

In this work, we propose \cmgan, a category-specific generative model for tileable SVBRDFs. Unlike previous methods, \cmgan (i) always produces \emph{tileable output}, (ii) enables more \emph{direct control} over the generated materials by optionally conditioning on input structure patterns that specify where the dominant features of the material should be located, and (iii) allows artifact-free reconstruction of structured materials from a \emph{single photograph}, due to the additional regularization provided by the input condition.

\cmgan is a variant of StyleGAN2~\cite{StyleGAN2} whose architecture is modified to always output tileable material maps. It optionally uses features extracted from a conditional input pattern in addition to the standard ``style'' latent code. Fixing the conditional pattern while varying the style code allows generating materials with the same dominant features, but different styles.
%a fixed conditional pattern can be used to generate different styles with different latent codes, all of which follow the condition. %\KS{flagging this since we are making a strong disentanglement claim here.}
We demonstrate that our conditional method can control the structure of complex materials like bricks and tiles, or the locations of wrinkles in leather.
%The semantics of the patterns differ per material category: for tile or stone materials, they specify the tile/stone locations and (optionally) average colors, while for leather and ground they may specify the main surface variation. Instead of using a single generator, we use multiple generators for different broad material categories; in this paper, we use tiled materials, leather materials and ground materials. 

Based on \cmgan, we propose an inverse rendering approach to find a material matching the style of a \emph{single} target photograph taken with flash (though our model can support other lighting or multiple inputs). For conditional materials, our method enables the user to provide an approximate pattern corresponding to the target photo; the pattern need not be precisely aligned with the target image. %Optimizing the parameters of \cmgan (the latent code and noise) to best explain the target photograph produces high-quality reconstructions. 
Unlike previous per-pixel approaches, the resulting materials are tileable, can be larger than the target image, and are controllable by varying the condition pattern. These additional benefits from \cmgan could enable artists to easily generate and fit a wide variety of material variations to reference photographs and will become integral to practical material authoring workflows. %\vde{Not sure the last sentence is needed here.}

%% file: related.tex
\section{Related work}

\paragraph{Material acquisition} 
Recovering material properties from images is a long-standing challenge in graphics~\cite{Guarnera2016}.
Deep learning demonstrated remarkable progress in the quality of SVBRDF estimates from single images (usually captured under flash illumination) \cite{Li2017,Deschaintre2018,Li2018}. These approaches produce smooth, plausible maps, but still suffer from artifacts caused by over-exposure burn-in, despite recent architecture modifications~\cite{Guo2021} or GAN-based, mixed data augmentation~\cite{Zhou2021}. Additionally, a single input is often too under-constrained, leading to inaccurate specular material properties. Recently, Henzler et al.~\cite{Henzler2021} proposed to leverage a stationarity prior to capture SVBRDFs from a single flash photograph, reducing burn-in artifacts but limiting the class of supported materials.
To improve quality, few-images methods were proposed through direct inference~\cite{Deschaintre2019, Ye2021} or deep optimization ~\cite{Gao2019, Guo2020}. %The optimization-based approach allows Gao et al.~\shortcite{Gao2019} to ``check'' the results at runtime and improve them. They propose to directly optimize the latent space of a learned material auto-encoder to minimize the rendering error of the output material maps against the input images.
Specifically, MaterialGAN~\cite{Guo2020} proposes to train an unconditional generative model for materials, allowing to optimize in the generator latent space and input noise to match the appearance of a target given a few target pictures. However, this model entangles spatial layout and style, is not tileable, and gives no explicit control over the generated materials.  
%Sampled materials do not always respect their class semantics; for example, tiles can be randomly rotated compared to the canonical orientation. 

%However, the results are not tileable nor explicitly editable. Through our structure-aware conditional models our approach enables pattern and style control and can create tileable results from non-tileable target photos, a crucial property for practical applications.

%Our key insight---conditioning material generation on a spatial pattern---makes our model structure-aware and enables controllability. Our approach is able to produce tileable results from non-tileable target photos, which is crucial for practical applications but not possible with the original MaterialGAN.
%and is trained on a single large set of varied materials, not per material category, which limits available control. 

As opposed to most of these approaches, we do not target exact per-pixel capture, but rather a result with matching perceptual style, guided by the photo. Our approach enables guided recovery of tileable and meaningfully controllable materials from a single picture and an optional structural pattern, through category-specific conditional GANs. 

Several works explicitly explored tileability for texture synthesis~\cite{pmlr-v70-bergmann17a} or completing textures into tiles through feature repetition and tile search~\cite{rodriguezpardo2022SeamlessGAN}. Our approach is to instead enforce tileability through our network architecture; we do not attempt to complete non-tileable textures into tileable ones. Instead, we define a space of plausible tileable materials, which we can randomly sample, or project target photographs into it.

\paragraph{Guided material generation and acquisition}
To provide control in the material acquisition and generation process, recent work proposed to leverage different guides. Using one~\cite{Deschaintre2020} or multiple images~\cite{rodriguezpardo2021} alongside a small material exemplar, previous work propose to transfer properties to large scale photograph of similar materials. Guehl et al.~\shortcite{Guehl2020} define a procedural generator for realistic textural structure masks, which they use to propagate existing material properties. Leveraging this procedural structure generator, Hu et al.~\shortcite{Hu2022} recently proposed to generate an entire procedural material graph using a segmentation mask and material as input. As opposed to our approach, these methods rely on pre-existing material inputs; they do not target material generation or capture.%, but rather the transfer of known material parameters to similar areas.

Procedural material representations in the form of node graphs \cite{Substance}, represent an artist-defined set of material appearances. Hu et al.~\shortcite{Hu2019} and Shi et al.~\shortcite{Shi2020} propose to optimize the parameters of existing material graphs to match the appearance of a target photo. Our method is related to procedural materials in that it establishes a function from an input pattern (a generator in the case of procedural materials) to an SVBRDF. However, it is more general than these approaches, as it trains generators for broad material classes, where training data can come from any source. \cmgan could be considered a class-level, neural procedural representation, capable of generating a large variety of samples, with conditions on the local structures and style.

\paragraph{Generative models}
Our method is based on the StyleGAN~\cite{StyleGAN,StyleGAN2} architecture which has demonstrated compelling image synthesis results on well defined domains such as faces, landscapes, or cars. While recent approaches \cite{CIPS} propose independent per-pixel generation through a unique latent vector and positional encodings, most works leverage spatial convolutions. 
Isola et al.~\shortcite{pix2pix} proposed an image-to-image translation model that conditions generation on an input image. More recently, different methods such as SPADE~\cite{spade} and CollageGAN~\cite{CollageGAN} proposed to condition GAN generation with semantic masks. We build on this work, conditioning the generation of materials with structure through a binary or color mask. %, which carries different meanings, depending on the generated material class.
Recently, Karras et al.~\shortcite{StyleGAN3} highlighted the importance of careful signal processing and translation invariance in GANs to prevent the network from building priors based on absolute position in the image. We prevent dependence on absolute positions by our tileable architecture modifications and by random translations of our dataset and generated materials. %Perfect translational invariance with respect to input pattern translation is not required for our application, though it would likely be achievable by carefully designing tileable versions of the low-pass filtering operations developed in StyleGAN3.

%\begin{itemize}
%#    \item Deschaintre 20: material example as a guide to texture image
%    \item Hu 19, MATCH: existing procedural graph models (akin to our trained GAN)
%    \item PPTBF: Focus on generating mask, given a mask matching a material, they can propagate this material to new masks but they don't generate anything new
%    \item Hu 22: Given a material, produce mask, procedural representation for mask and material and optimize this. It requires user input and a material, it doesn't do the acquisition, it is a bit slow and not tileable (A bit weak differences, but we should showcase it). 
%\end{itemize}
%Deschaintre 20, Hu 22, MATch, Hu 19, PPTBF
%Introduce what is meant by guided material generation, why it's interesting and the diversity it allows 
%Guehl et al. \shortcite{Guehl2020} introduce TODO.

%% file: forward.tex
\section{Tileable and Conditional GAN}
\label{ssec:gan}

In this section, we describe the architecture of \cmgan, as well as details about training techniques and training data.
%\cmgan is based on StyleGAN2 \cite{StyleGAN2}.%, which is an improvement of StyleGAN \cite{StyleGAN}. While other architectures have been recently successful in the GAN space \cite{StyleGAN3,CIPS}, StyleGAN2 is still heavily used, both directly and as a basis for derived methods.

\begin{figure}[t]
    \centering
    \includegraphics[width=1\linewidth]{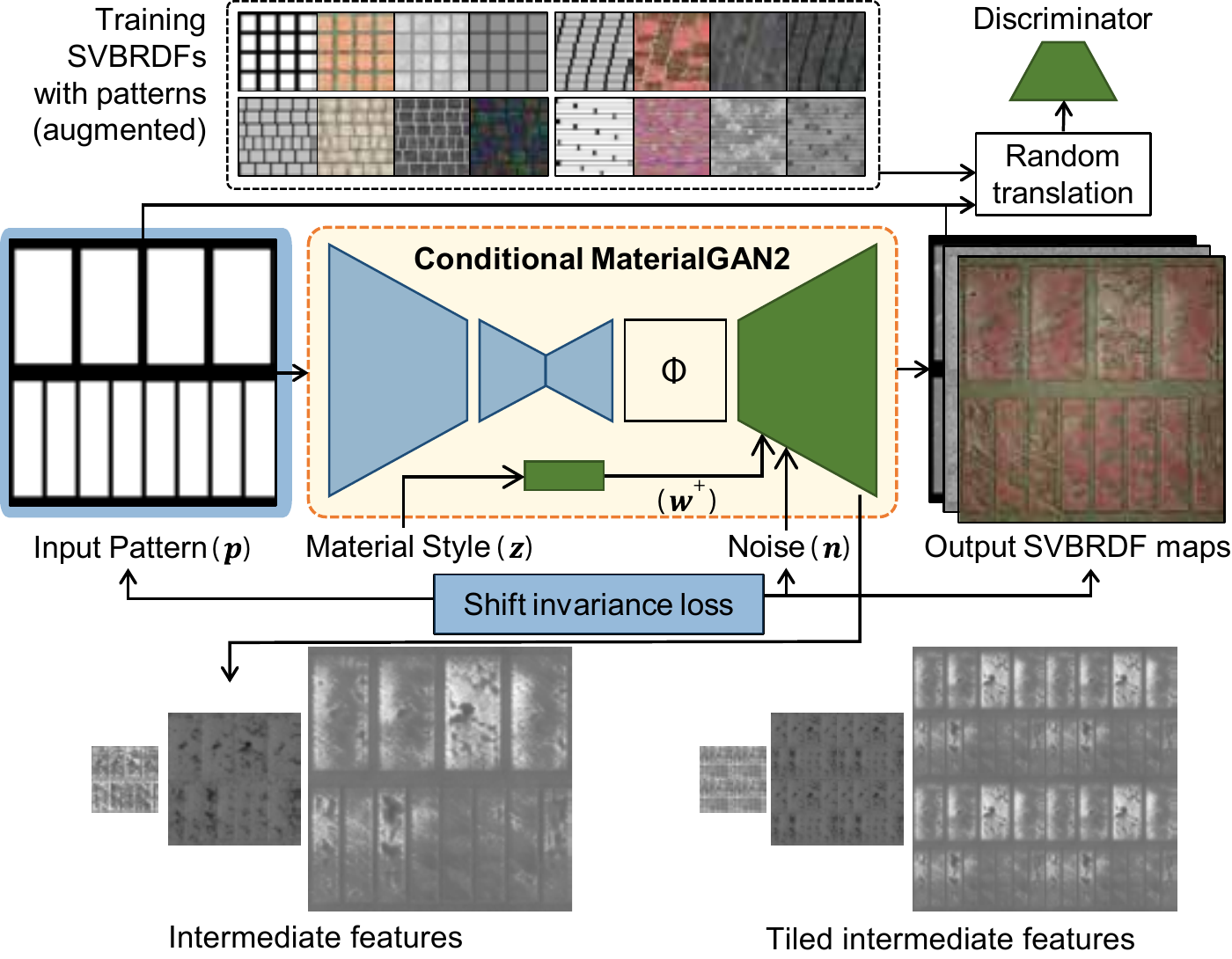}
\caption{The conditional version of \cmgan is trained on a dataset of SVBRDF parameter maps with corresponding condition patterns. Our conditional generator has a CollageGAN-like encoder that maps the input pattern $\pat$ into features $\bm{\phi}$ at the start of the StyleGAN2-based decoder (green); the decoder also receives the latent code $\bz$ (via a mapping network) and noise. The encoder and decoder have been modified to only use tileable operations. The resulting SVBRDF maps, together with the condition, are randomly translated and fed to a StyleGAN2 discriminator. Differences between conditional model and unconditional model are shown in light blue. In unconditional model, we do not have input patterns and encoder-decoder. See Sec. \ref{ssec:gan} for more details.}
\label{fig:architecture}
\end{figure}

\begin{figure}[tb]
    \centering
    \includegraphics[width=1\linewidth]{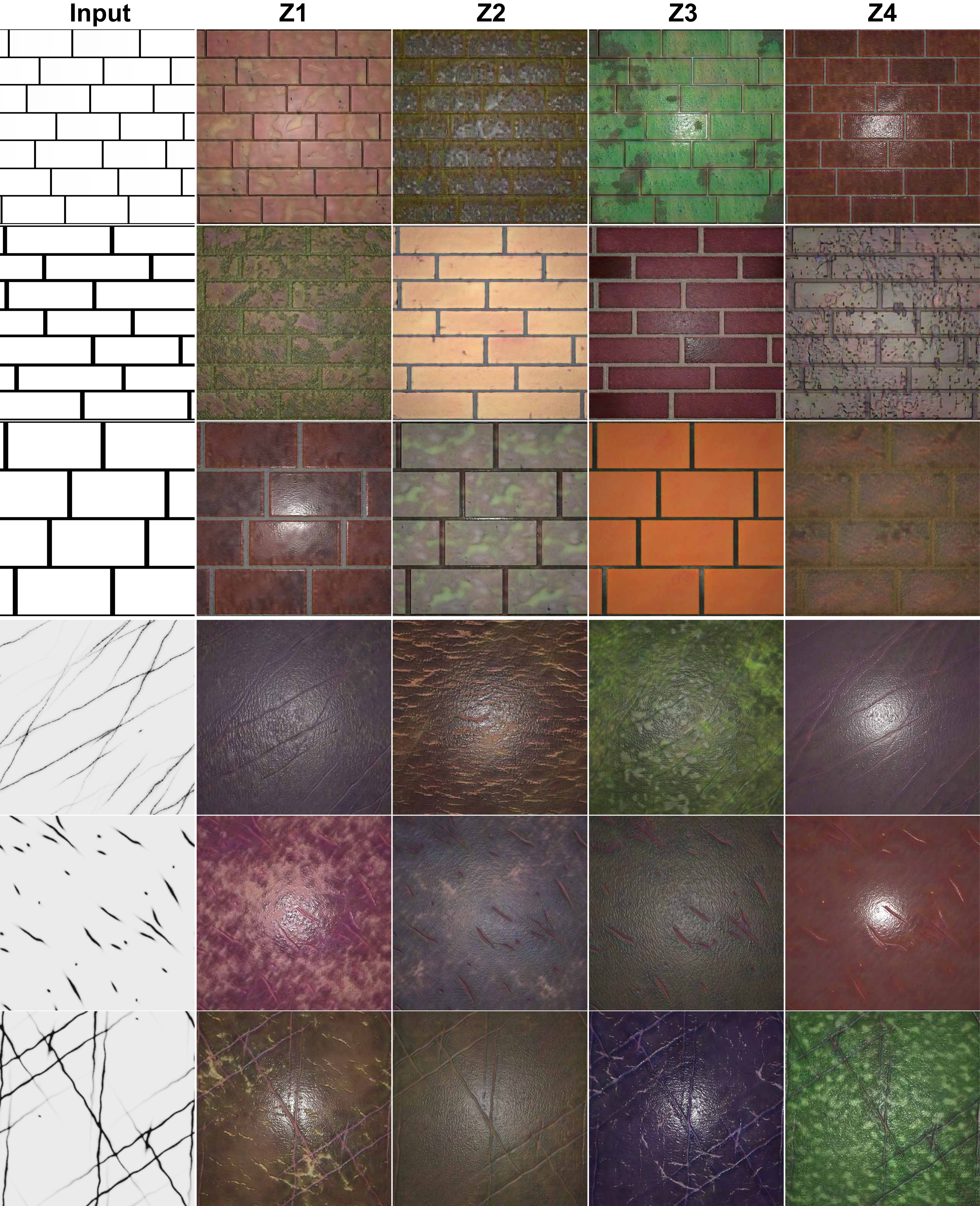}
\caption{Randomly sampled conditional materials from the tile and leather classes. We feed the conditional pattern on the left to \cmgan, along with four different random latent vectors $\bz_1,\dots,\bz_4$, to produce corresponding material instances. The results have varied appearance with a constant layout condition. The texture maps are shown in supplementary materials. }
\label{fig:rand_sample_cond}
\end{figure}

\begin{figure}[tb]
    \centering
    \includegraphics[width=1\linewidth]{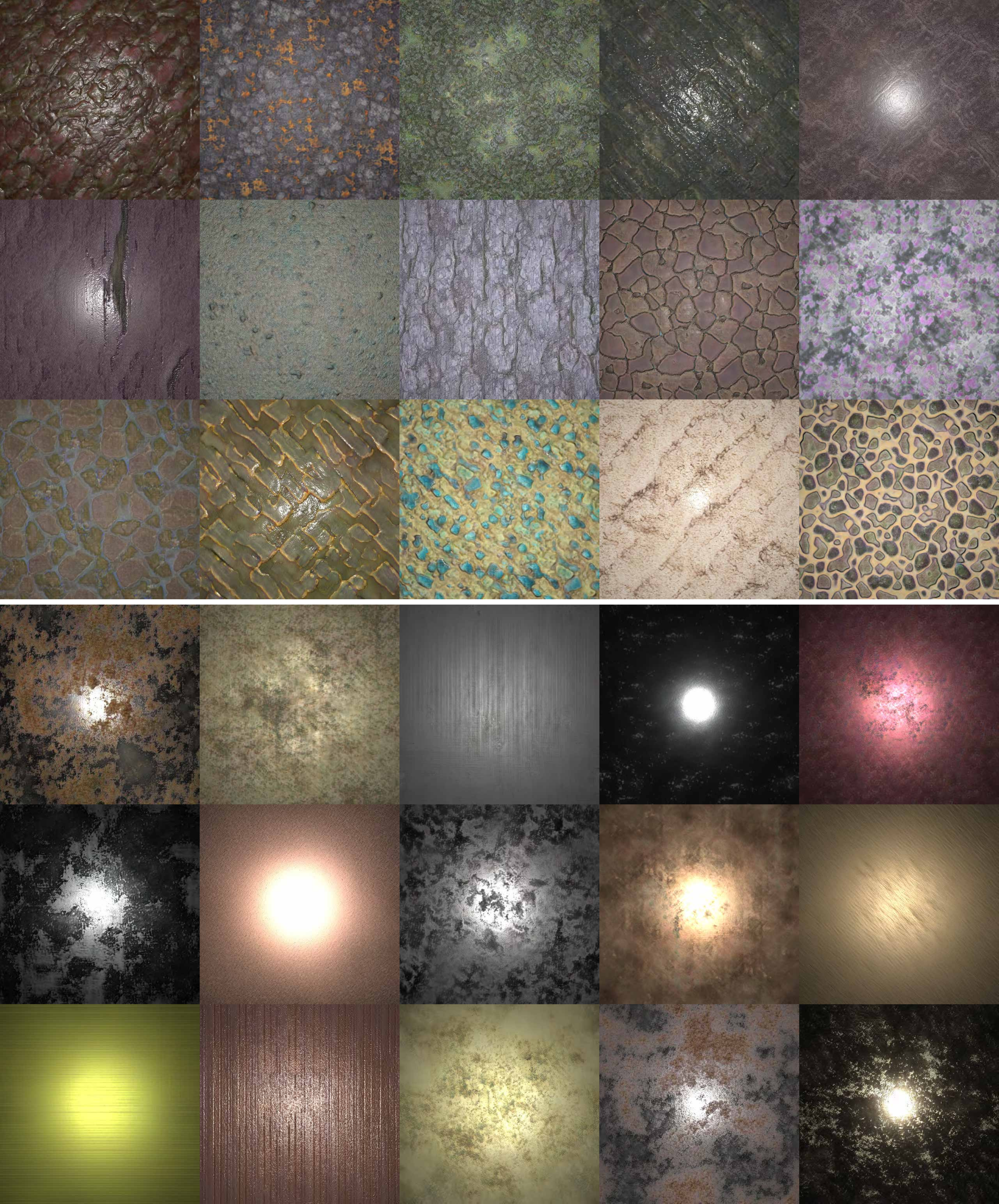}
\caption{Randomly sampled unconditional materials from the stone and metal classes, showing the diversity of the results within each class. The corresponding texture maps are shown in supplementary materials.} %Figure S\ref{fig:uncond_supple1}. }
\label{fig:rand_sample}
\end{figure}

\subsection{Tileable architecture}
\label{ssec:tileable}
\cmgan is based on StyleGAN2~\cite{StyleGAN2}.
%An important feature of our architecture,
Different from StyleGAN2, our architecture is designed to preserve tilability, a property that is important for practically usable textures. 
%by only using operations that 
%or \emph{periodicity}:
%is its tileability or \emph{periodicity}:
%the model is designed to only generate periodic (tileable) signal, which is very important for practically usable textures.
This is achieved by replacing all convolution, upsampling and downsampling operations in the model by wrap-around versions, agnostic to their location in the image or to the boundary location. Therefore, these operations always produce smoothly tileable results
%that are agnostic to the boundary location and preserve continuity across the boundary.
%preserve periodicity.
%Therefore, none of the operations in the model have the ability to tell where in image space (relative to the boundary) they are located, and are forced to produce smoothly tileable results,
if the input condition is tileable, which is always true in our results.

We observe that in our modified architecture, intermediate feature maps at all levels remain periodic (shown in Fig.~\ref{fig:architecture}). %In fact, this is true from the beginning of training, rather than a learned property.
Furthermore, we found that this architecture can even be used to produce tileable materials from a non-tileable dataset (though the models shown in the paper use tileable datasets).

\subsection{Conditional architecture}
\label{ssec:conditional}
CollageGAN \cite{CollageGAN} is a recent conditional GAN approach for generating realistic images given segmentation maps. Though we do not currently collage multiple GAN models, we use a conditional architecture (shown in Fig. \ref{fig:architecture}) inspired by CollageGAN.%, which in turn is derived from StyleGAN2, with some modifications.

The main idea of the CollageGAN architecture is that the conditional pattern $\pat$ is encoded into features $\bm{\phi}$, which feed into the initial layer of the architecture at size $32 \times 32$, replacing the learned initial constant of StyleGAN2. This approach is also used in our architecture; however, a major difference from CollageGAN is that our latent vector $\bz$ is generated at random from the normal distribution, instead of depending on $\pat$. We therefore do not apply any KL-divergence loss to regularize the latent vector either. This provides a certain level of disentanglement between the pattern $\pat$ and the ``style'' of the material, given by the latent vector $\bz$. This property is important for our inverse rendering, and does not exist in CollageGAN, where the result depends deterministically on the condition $\pat$.

Our trained generator can be written as a function $\generator$ that generates the material parameter maps (in our case, albedo $\albedo$, height $\height$, roughness $\rough$), given the latent code $\bz$ and pattern $\pat$. For our metal material class, the generator also outputs the metallic amount $\mathbf{m}$; in general, it could output any number of channels driving any shading model.

\subsection{Synthetic dataset design}
\label{ssec:dataset}
%
%We design a dataset for each material category---specifically, \KS{tile, leather and ground}---to train each version of \cmgan. 

We trained four \cmgan models using synthetic datasets: unconditional stone and metal models, and conditional tile/brick and leather models. For a given class, we collected a number of Adobe Substance node graphs~\cite{Substance}. We sampled their parameters in reasonable ranges, and saved the output material maps: diffuse albedo $\albedo$, height (displacement) $\height$, roughness $\rough$, and metallicity $\mathbf{m}$ for metals (though other material categories such as fabrics or fur could output other maps). We further augment the training examples, by applying augmentation to the height and roughness channels. We apply small color augmentation to remain within realistic color tones for a given material class.

Pairing each synthetic material with its conditional pattern is dependent on the specific material category. For the tile/brick class, we identified a node in the corresponding graph whose output approximately matches our desired pattern, having a high value denoting a tile and a low value denoting the gap between tiles. For the leather class, we similarly identified a node whose output closely matches the final wrinkle pattern. We either threshold these outputs to make them binary for tile or keep these outputs as grayscale for leather, yielding the pattern $\pat$, which we save as an additional texture output. We kept the stone and metal categories unconditional, as there is no obvious structure shared by them (though specific subsets of stones or metals could have structure and we could make specific models for them). %\KS{Might be useful to show some training set examples of patterns + materials.} 

%For the other two categories (patterned leather and ground), we used a simpler approach of taking a blurred version of the heightfield as the condition, adjusting its range and contrast to be approximately consistent between material instances. Thus, the condition specifies the rough shape of the material heightfield.

\subsection{Losses}

We train our networks with an adversarial loss $L_{\text{adv}}$. We follow the StyleGAN2~\shortcite{StyleGAN2} discriminator architecture and regularization, except with more channels provided to the discriminator, representing our parameter texture maps.

For the conditional models, we also feed the condition pattern to the discriminator and randomly translate the pattern and maps (both ground-truth and generated) before feeding them. When training conditional models with only the adversarial loss above, we find that the model bakes in some material structure variations into the latent code $\bz$ (i.e., varying $\bz$ also changes the dominant structures in the generated material). However, we want the input pattern to drive the material structure; we introduce a second loss $L_{\text{shift}}$ ensuring that shifting the input pattern also shifts the generated material maps by the same amount. We run the pattern $\pat$ through the generator $\generator$ twice: the second time with a random translation (shift) $T$, maintaining the same latent vector $\bz$. A further detail is that we also shift the random noise (used in StyleGAN2 as additional inputs to the generator) by the same amount. Finally, the shift loss is defined as:
%\begin{equation}
$L_{\text{shift}} = \|T(\generator(\pat, \bz, \noise)) - \generator(T(\pat), \bz, T(\noise))\|_1$.
%\end{equation}
% This loss requires two forward passes of the generator but significantly helps the generated material to follow the structure given by the condition.

%We then undo this relative translation for the resulting texture, and encourage them to be equivalent. This additional loss adds extra cost (since the patterns are generated twice for the same latent vector), but we found it helps encourage condition translation invariance and disentanglement of style from condition.
% \subsection{loss function}
% \label{ssec:loss}

% our loss consists of two parts:

% \begin{itemize}
%     \item GAN loss: adversarial loss and regularization same as StyleGAN2 ~\cite{StyleGAN2}
%     \item Style regularize loss: 
%     \item shifted noise as the input to the StyleGAN2 decoder at different scale level 
% \end{itemize}

%% file: inverse.tex
\section{Matching target images} \label{sec:inverse}

\begin{figure*}[tb]
    \centering
    \includegraphics[width=\linewidth]{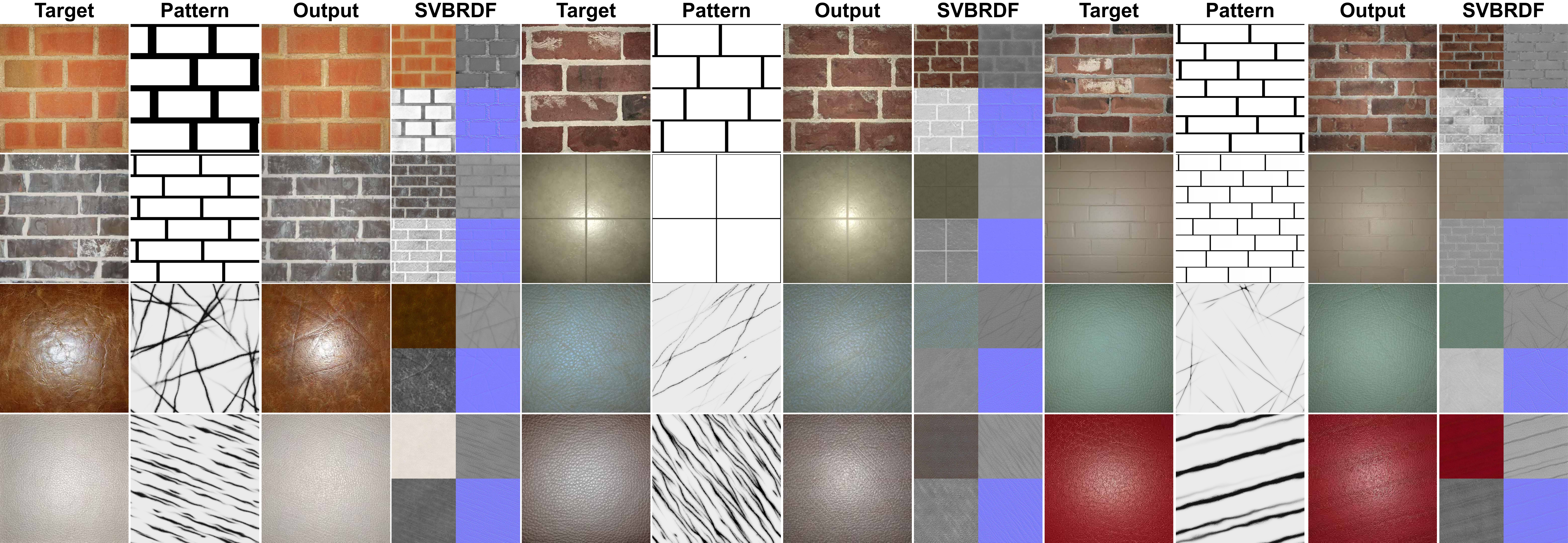}
\caption{Examples of reconstructing conditional materials from a \emph{single} target photograph with flash illumination for tiles (two top rows) and leather (two bottom rows). For each materials class, the target image is on the left, followed by the input pattern created by the user. For tiles we assume the user pattern has approximately matching feature sizes to the photo, though the feature layout can be arbitrary as we do not seek per-pixel correspondence. For leathers we assume the patterns represents wrinkles, though these wrinkles may not exist in the target images. The third column shows our resulting rendered material, followed by the predicted parameter maps (top-left: diffuse map, top-right:height map, bottom-left: roughness map, bottom-right:normal map). }
\label{fig:inverse_cond}
\end{figure*}

\begin{figure*}[tb]
    \centering
    \includegraphics[width=\linewidth]{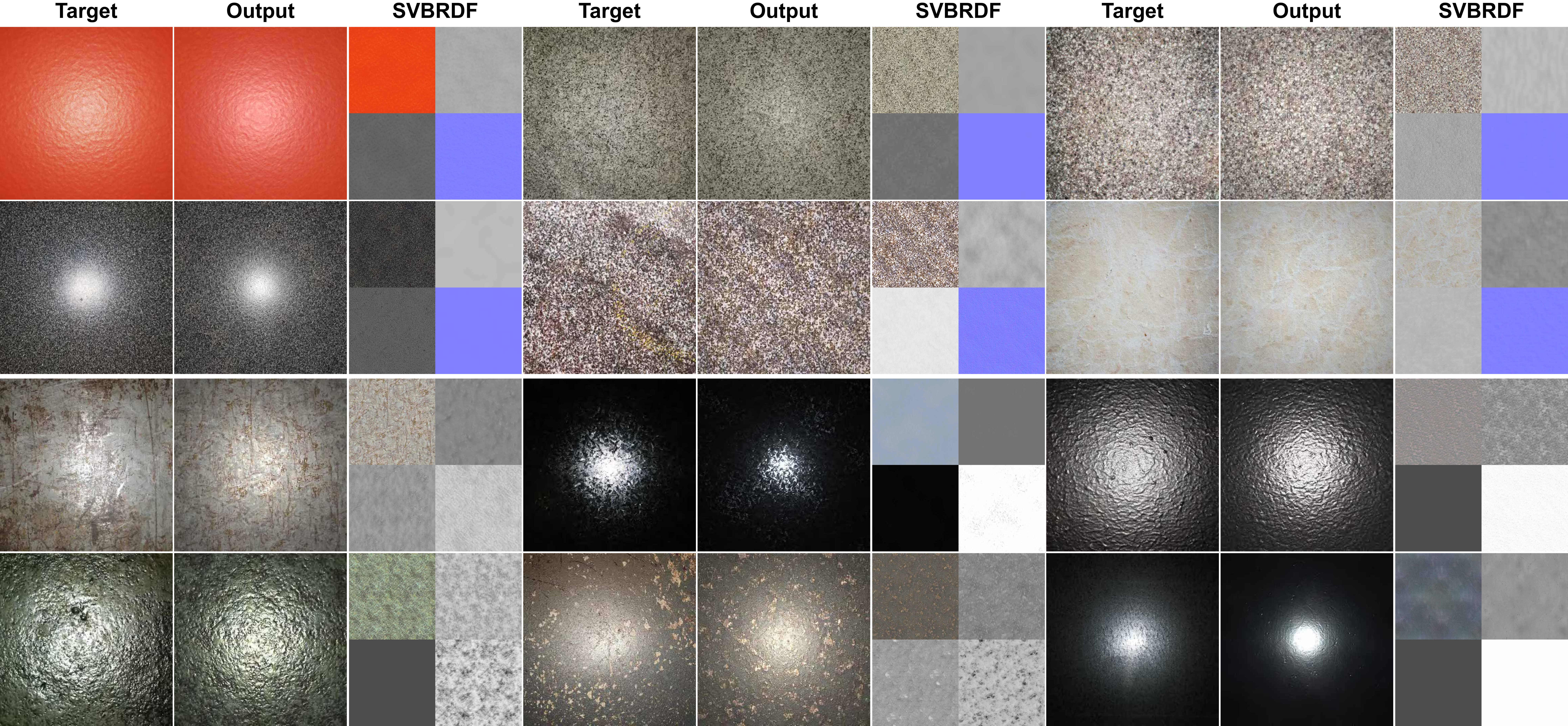}
\caption{Examples of reconstructing unconditional materials from a \emph{single} target photograph with flash illumination for stone (two top rows) and metal (two bottom rows). For each materials class, the target image is on the left, followed by our resulting rendered material and the predicted parameter maps (for stone, the order of feature maps is same as Figure \ref{fig:inverse_cond}; for metal, top-left: base color, top-right: height map, bottom-left: roughness, bottom-right: metallic map). }
\label{fig:inverse_uncond}
\end{figure*}

In this section, we use \cmgan to solve the inverse problem of finding an SVBRDF that, when rendered, matches the appearance of a target photograph of a physical material sample.

We assume the input image is a single photograph of a planar sample taken with a cell phone camera with flash. We use a differentiable rendering operator $\render$ that takes as input the parameter maps, and synthesizes corresponding images of the material lit by the flash illumination. We turn the height map into a normal map using central finite differences, and shade the diffuse component using a Lambertian term, and the specular component using a standard microfacet BRDF with the GGX normal distribution \cite{Walter07}. This could be easily extended to other lighting/shading setups or to multiple inputs, as these are independent of \cmgan and are just modifications to the rendering operator.

Given a target image $\img$, we would like to find material maps that render to an image similar to $\img$, under a suitable loss $\loss$. More specifically, we define the vector $\latent$ to be the pair of style vector $\bw^+$ and noise vector $\noise$ in the $\calW^+ \calN$ space (illustrated in Fig.~\ref{fig:architecture}), instead of the original random code $\bz$, as detailed by Guo et al. \shortcite{Guo2020}. 
The optimization problem then becomes:
\begin{equation}
	\label{eq:opt_gan}
	\latentopt = \textstyle\argmin_{\latent} \loss(\render(\generator(\latent, \pat)), \img),
\end{equation}
where $\generator$ is the learned conditional \cmgan generator. (For unconditional generators, we drop the $\pat$ dependence.)
Given that both $\generator$ and $\render$ are differentiable operations, Eq.~\eqref{eq:opt_gan} can be optimized via gradient-based methods to estimate $\latentopt$ and the corresponding SVBRDF maps $\generator(\latentopt, \pat)$.

A key question is which loss $\loss$ to use. Per-pixel losses (as used by MaterialGAN) commonly cause overfitting, especially when run on single images. This leads to unrealistic maps with undesirable artifacts, such as flash leaking into the albedo, that happen to lead to lower loss than any realistic maps would~\cite{Deschaintre2018, Gao2019}. Our main loss term is a style loss based on the Gram matrix~\cite{Gatys2015} of VGG features~\cite{VGG}, which approximates the difference between distributions of neural feature activations, and is insensitive to pixel alignment. This loss has been successfully used for material capture without perfect pixel correspondence for procedural materials \cite{Shi2020,Guo2020bayesian}. The style loss drives the overall appearance, while our condition and prior learned by our generator ensures the local coherence and tileability of the results, and that the optimized appearance remains in the subspace of realistic materials for a targeted class.

We combine the Gram matrix loss with an $L_1$ loss computed between downsampled images of resolution $16 \times 16$. This improves global matching of overall color and roughness. Furthermore, at every other iteration, we apply random translations to the generated material maps before passing them to the rendering operator: essentially, we are looking for a material that renders to a close perceptual match to the target image across all of its translations. This further makes any remaining overfitting negligible. 

In contrast, the original MaterialGAN cannot be based on a global loss, because its material prior is not strong enough by itself. Using only a style loss and the low-resolution $L_1$ loss with our models works because they are class-specific and (in the case of highly structured materials like tiles) conditional on structure patterns. Tileability of the resulting materials is also only possible when using a style loss; a per-pixel match to the original (non-tileable) target image would of course not be able to produce tileable results. % In other words, our category-specific and conditional properties significantly regularizes the model spatially and allows us to optimize with just the style loss, which in turn enables us to obtain tileable results free from highlight leaks from a single photograph.

%% file: results.tex
\section{Results} \label{sec:results}

\begin{figure}[tb]
    \centering
    \includegraphics[width=\linewidth]{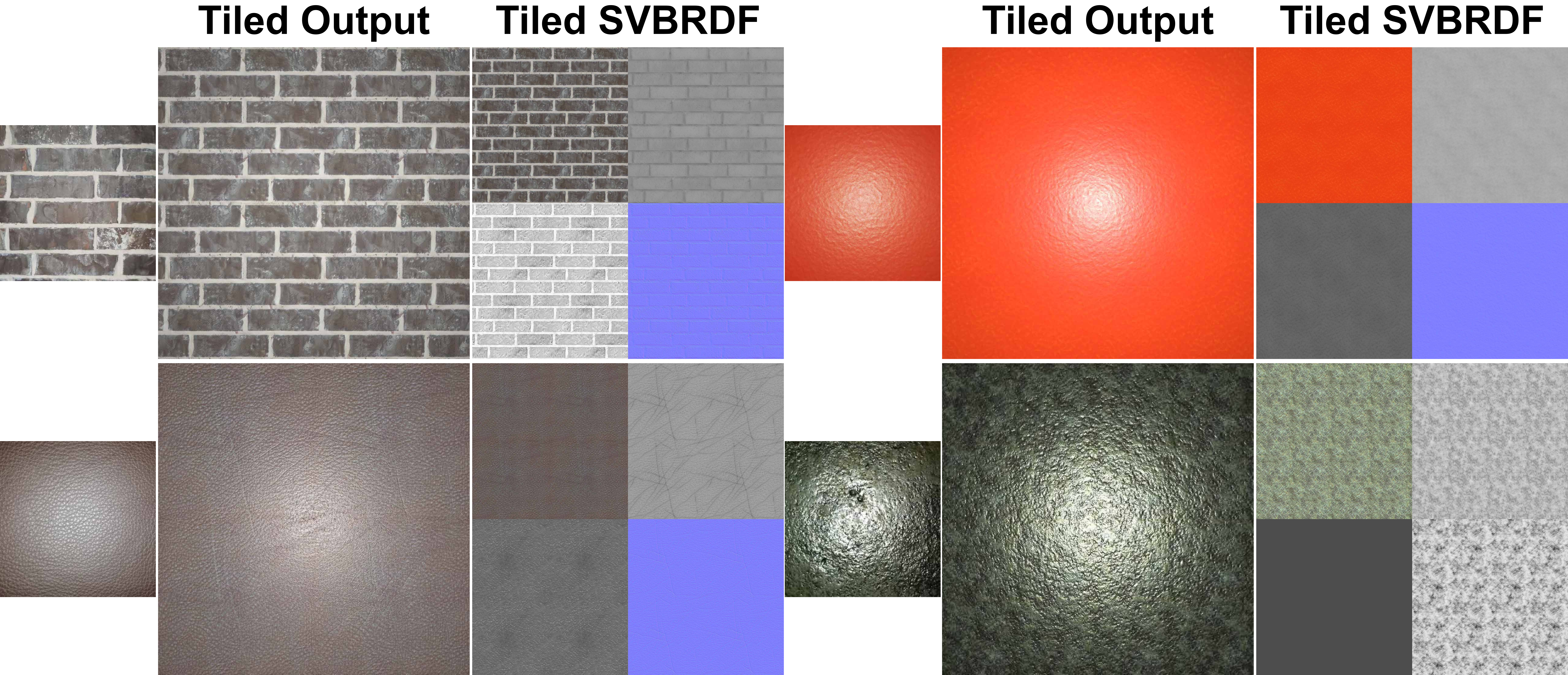}
    \includegraphics[width=\linewidth]{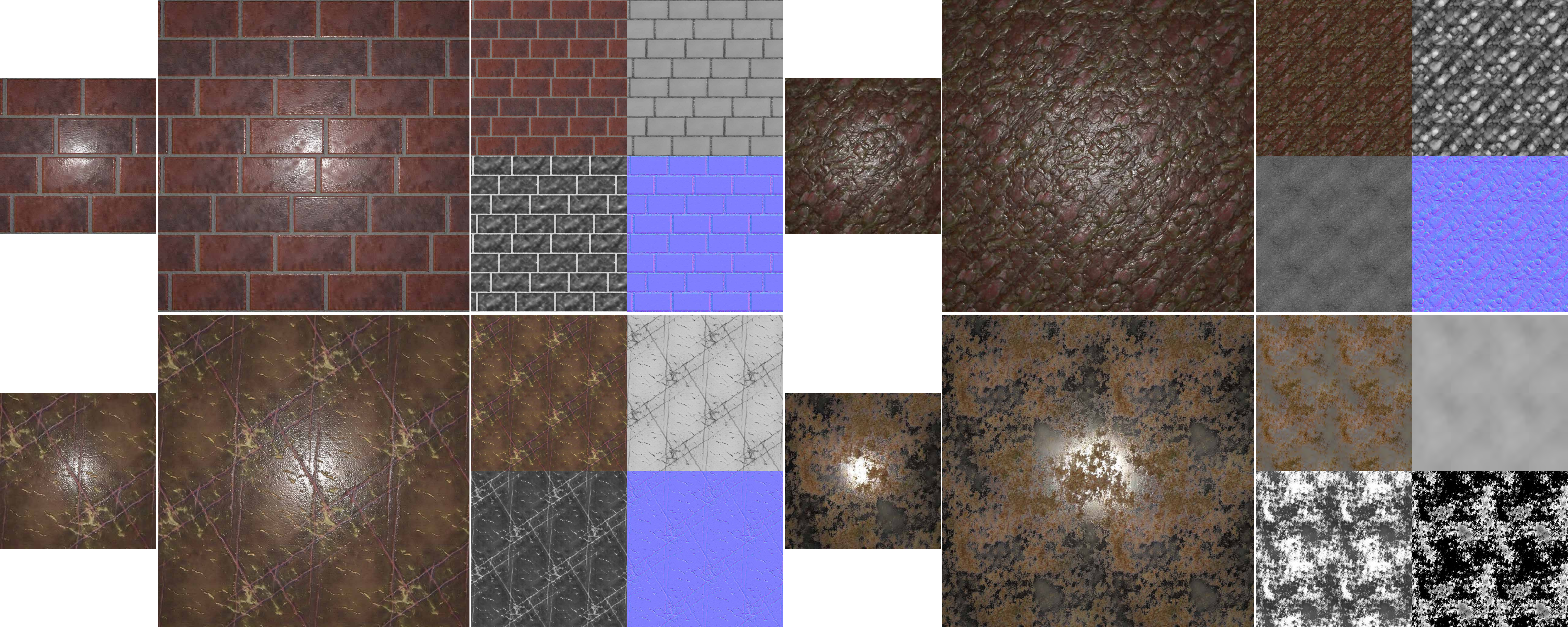}
\caption{ Demonstration of tileability for both inverse rendering (two top rows) and randomly sampled results (two bottom rows). The leftmost image is either the target for inverse rendering or the original generated material for randomly sampling, followed by a rerendered image using tiled texture maps and the corresponding tiled texture maps. The results show seamless tileability (periodicity) of our resulting textures, even though the target image is not tileable in the inverse rendering examples. }
\label{fig:tileability}
\end{figure}

\begin{figure}[tb]
    \centering
    \includegraphics[width=1\linewidth]{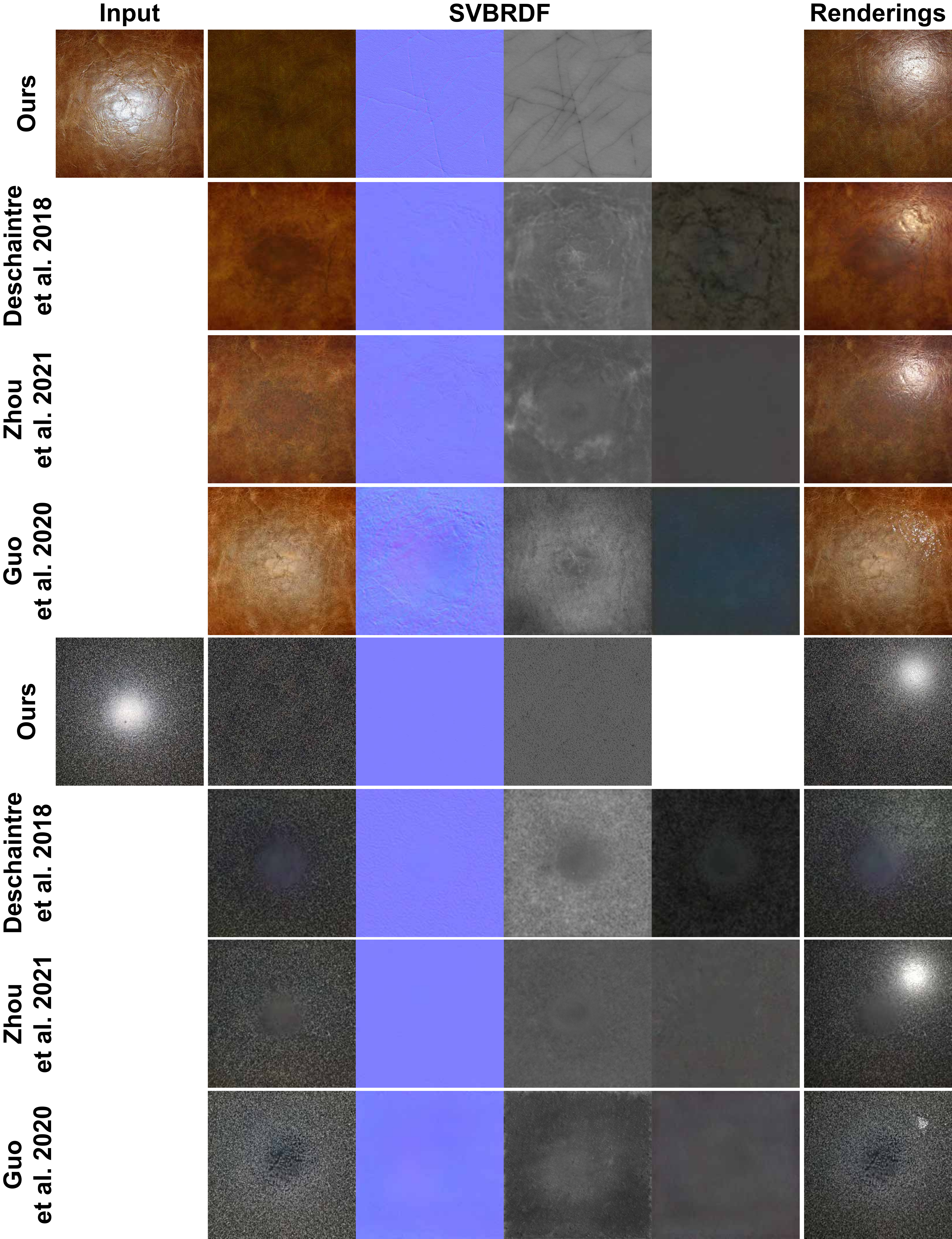}
\caption{Comparison with three SVBRDF estimation approaches~\cite{Deschaintre2018,Zhou2021,Guo2020} on SVBRDF capture from a single target image (left). All of these approaches generate unclean feature maps, baking in the flash highlight. In contrast, our material maps and re-renderings are clean and plausible.}
\label{fig:Comparison}
\vspace{-0.5cm}
\end{figure}

\begin{figure}[tb]
    \centering
    \includegraphics[width=\linewidth]{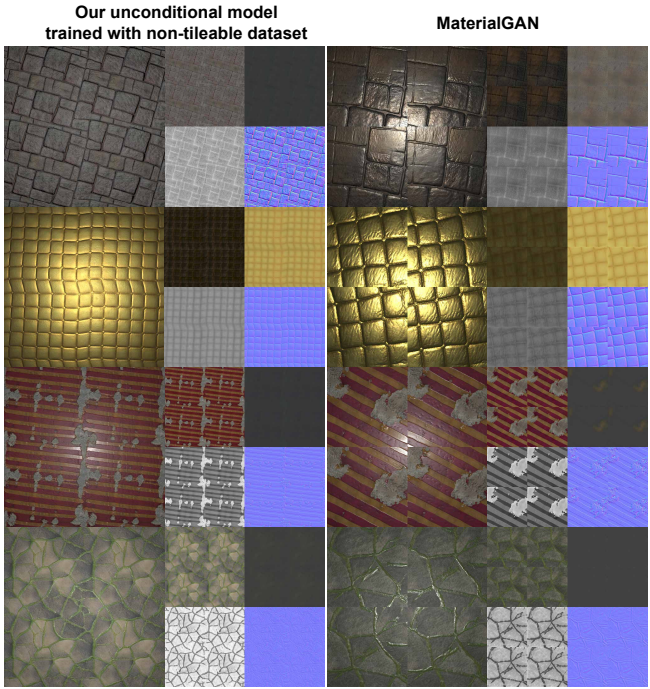}
\caption{ Comparison of randomly sampled results of original MaterialGAN and our unconditional model trained on the same non-tileable dataset MaterialGAN was trained on, showing 2x2 tiled results and the corresponding tiled material maps. Even when trained with a non-tileable dataset, our unconditional model can produce seamless texture maps.}
\label{fig:Comparison_MG_sample}
\end{figure}

\begin{figure}[tb]
    \centering
    \includegraphics[width=1\linewidth]{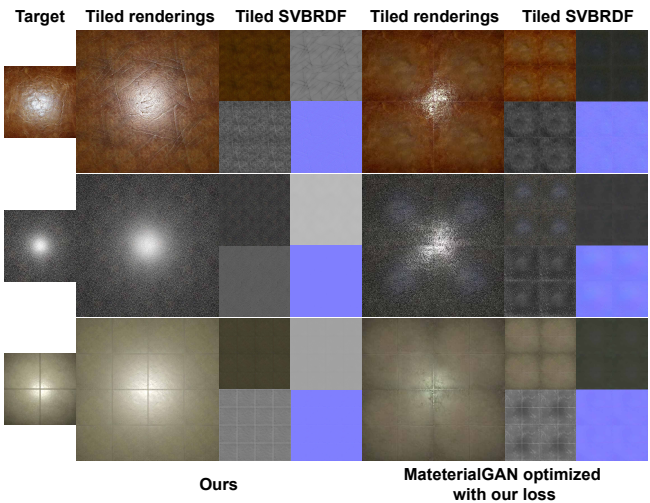}
\caption{Comparison to the original MaterialGAN optimized with our global loss. Even when using a non pixel-wise loss, the results of MaterialGAN present highlight baking artifacts and are not tileable, unlike our results.}
\label{fig:Comparison_MG_inverse}
\end{figure}

We trained several versions of
\cmgan: conditional models for brick/tile and leather materials, and unconditional models for stone and metal materials. We additionally trained a color-conditioned version shown in supplemental materials. Here, we demonstrate the results of applying these models to both forward generation and inverse optimization tasks, with and without conditions. We show more results in supplemental materials.

\paragraph{Forward generation}
We demonstrate the results of using our models in a forward manner for material generation. In Figure \ref{fig:rand_sample_cond}, we show the results of several input patterns provided to our conditional generator, with four randomly sampled styles (latent codes) for each pattern, and showcase the quality and variety of the generated results.
This demonstrates that the layout of the results indeed follows the input condition, while the style of the result is separated and depends on the latent style code. For conditional models, all input conditions are tileable (periodic). Such patterns are easy to create using existing tools such as Substance Designer's generator nodes. In Figure~\ref{fig:rand_sample}, we show the results of our unconditional generators, showing semantically meaningful variety as we sample new materials.

\paragraph{Material capture} 
Next, we show examples of using our generative models as priors for optimization-based material capture from real photographs. In Figure \ref{fig:inverse_cond}, we show conditional material capture from target images. We only assume that the input patterns match the approximate feature size and layout in the target image, with no pixel correspondence required. Note how our reconstructions capture the appearance of the material samples while remaining faithful to the input condition, even though it is not exactly aligned with the target image. We also show that we can change the conditional patterns, allowing to control the result details while preserving the target appearance. Next, in Figure~\ref{fig:inverse_uncond} we show examples of materials recovered using our unconditional models: stone (2 top rows) and metal (2 bottom rows). By using a class-specific generative model, we can relax the optimization loss from per-pixel to global, preventing the flash artifacts typical in previous single image acquisition methods, and obtaining tileable results as an additional benefit.

We evaluate further properties of our method: the tileability of our results (Figure \ref{fig:tileability}). The supplementary materials also demonstrate that the effect of input patterns on results, the invariance to translation of the input pattern, the use of a colored condition, using conditions of higher resolutions.

\paragraph{Comparison} We first compare our approach to MaterialGAN \cite{Guo2020}, Zhou et al.~\shortcite{Zhou2021} and Deschaintre et al.~\shortcite{Deschaintre2018} in Figure~\ref{fig:Comparison}. We see that the MaterialGAN optimization approach, when used with a single input photograph, does not sufficiently constrain the space of realistic materials and generates unrealistic material parameters. The forward methods by Zhou et al. and Deschaintre et al. also suffer from flash artifacts and do not recover tileable materials. 
Our method does not suffer from these problems even when matching a single target image, thanks to its architecture, the pattern conditioning, the use of a style loss rather than a per-pixel comparison, and the random translations during optimization. While our results are not exactly aligned with the target photograph, they capture the overall appearance more accurately and are significantly more practical in a content creation workflow.
We further compare our tileable architecture to MaterialGAN by training our unconditional model with the original dataset from Deschaintre et al. that MaterialGAN was trained on (Figure \ref{fig:Comparison_MG_sample}), showing that our results are tileable even if the dataset is not.
We also compare our inverse rendering results with the original MaterialGAN but using our loss function in Figure \ref{fig:Comparison_MG_inverse}, showing that just changing the loss is not sufficient to improve the results.

%As our method is trained with a semantic understanding, our network benefits from better priors to match the target appearance with realistic material parameters.

\subsection{Discussion and Limitations}
% vThe amount of control available through the condition is lower than the parameters of a fully procedural node graph. This could be partially addressed by including our generator itself in a node graph, and applying further nodes for minor adjustments to the input pattern and output maps. \vde{This limitations feels like we are shooting ourselves in the foot: it's obvious we will not match a complex node graph expressivity, the proposed solution sounds a bit fishy. Should we just remove this?} \milos{Agree}

Our approach is trained per material semantic class, which enables more meaningful style navigation and conditioning, but requires training one network per class and to have sufficient amount of ground truth data for each class we want to represent. Our current generators have a resolution of $512 \times 512$ and need to be trained for several days on 4 NVidia V100 GPUs. Higher-resolution generative models with faster training time remain a desirable future improvement. Our optimization takes around two minutes per target image, which is usable in practice, but a faster solver (perhaps based on a neural network predicting larger steps than basic gradient descent methods) would be a valuable direction. %A single flash image may not contain enough information about a material, regardless of the method used; other lighting configurations would be straightforward to support.

%\vde{While our method can handle varying material in a capture if semantically meaningful (ie grout and bricks), it does not support the acquisition of highly spatially varying material presenting multiple semantic classes in a single SVBRDF. Similarly our low resolution l1 loss in the inverse rendering context ensure a rough alignment of features, but doesn't guarantee an exact reproduction when it is perceptually important (e.g. flower patterns)}

%We train one GAN per class, it has pros (editability, semantic relevance) but also cons (need enough data in the class, multiple networks, ...).

%% file: conclusion.tex
\section{Conclusion and Future Work}
We propose a new material authoring approach conditioned on class-specific networks, and show that we can condition them on easily authored inputs such as structure for bricks or cracks for leather. We train a semantically meaningful material generator, separating style from structure in the available control. We demonstrate material generation, acquisition matching the appearance of a single photograph, as well as additional effects such as interpolation, on multiple material classes. 
We believe that our method simplifies material design, making it accessible to novice user with extended control, without relying on complex existing material graphs.

We hope that future work will increase the resolution and training performance of the conditional generators, while extending them to new categories. Training \cmgan on actual material sample photographs from specific categories is an exciting future direction, creating highly valuable material generators based on real data.

%% file: supple.tex
% \widetext
\clearpage

\pagebreak

% Title portion
\twocolumn[
\begin{center}
  \huge \textbf{Supplementary Materials for \cmgan: \\Tileable, Controllable Material Generation and Capture} 
\end{center}
\vskip 2em
]
% \maketitle

% \section{Additional results and ablations}

\begin{figure*}[tb]
    \centering
    \includegraphics[width=1\linewidth]{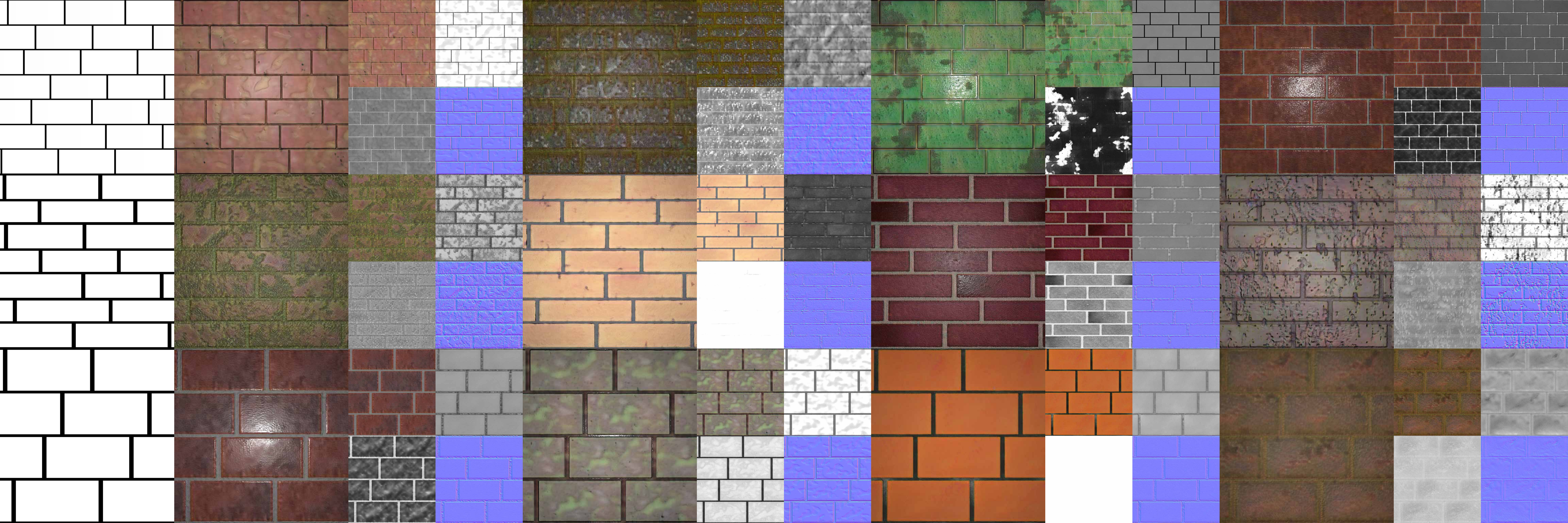}
    \includegraphics[width=1\linewidth]{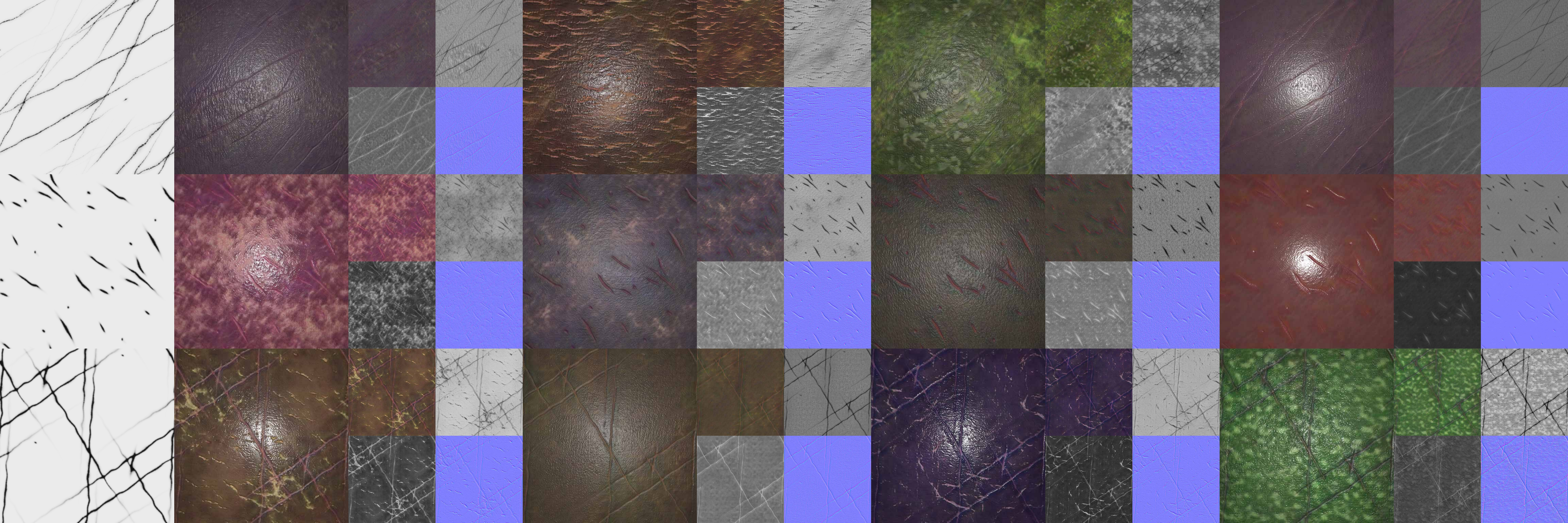}    
\caption{Randomly sampled conditional results with the generated texture maps of tile and leather examples in the Figure 3 of the paper.}
\label{fig:cond_supple1}
\end{figure*}

\begin{figure*}[tb]
    \centering
    \includegraphics[width=1\linewidth]{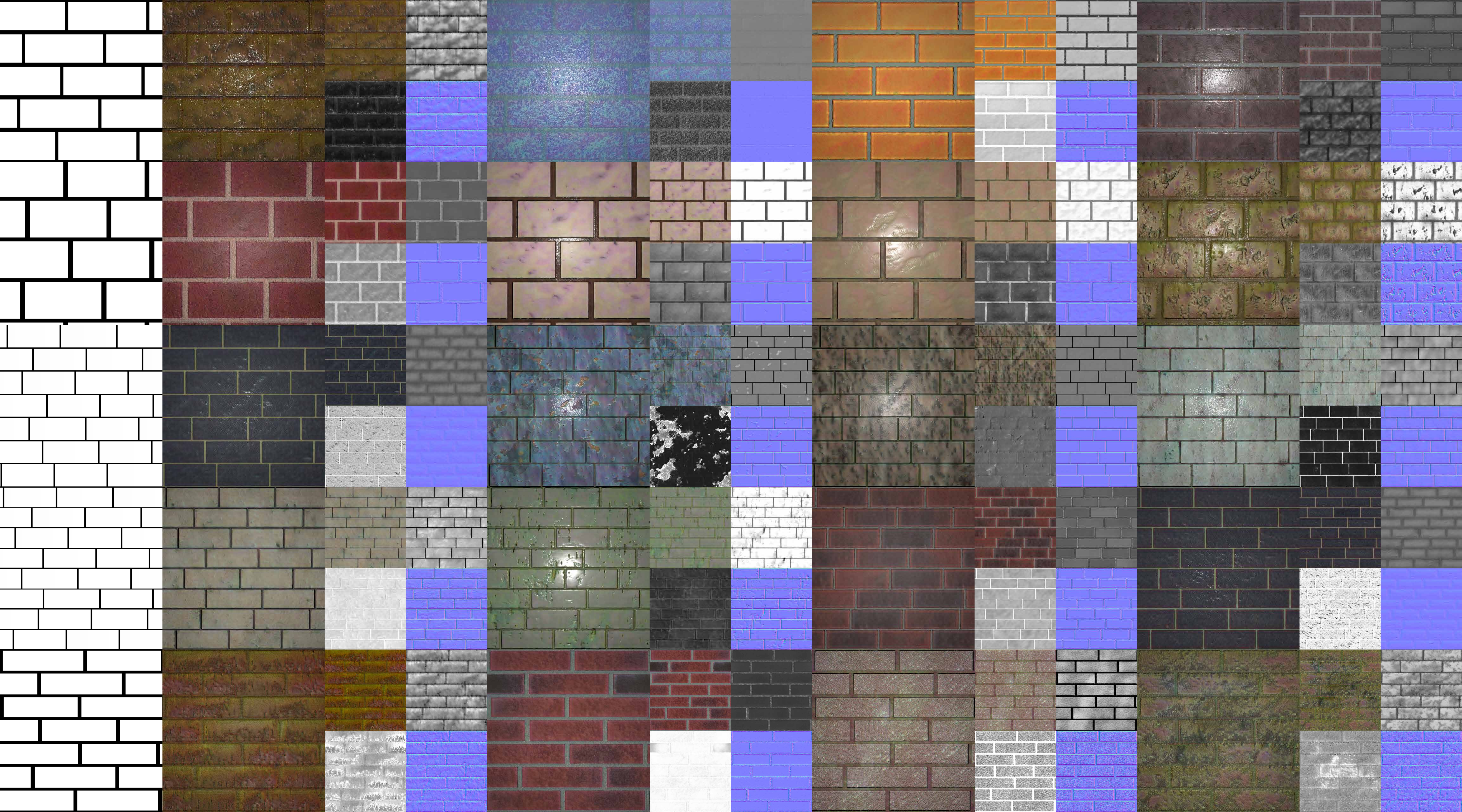}
    \includegraphics[width=1\linewidth]{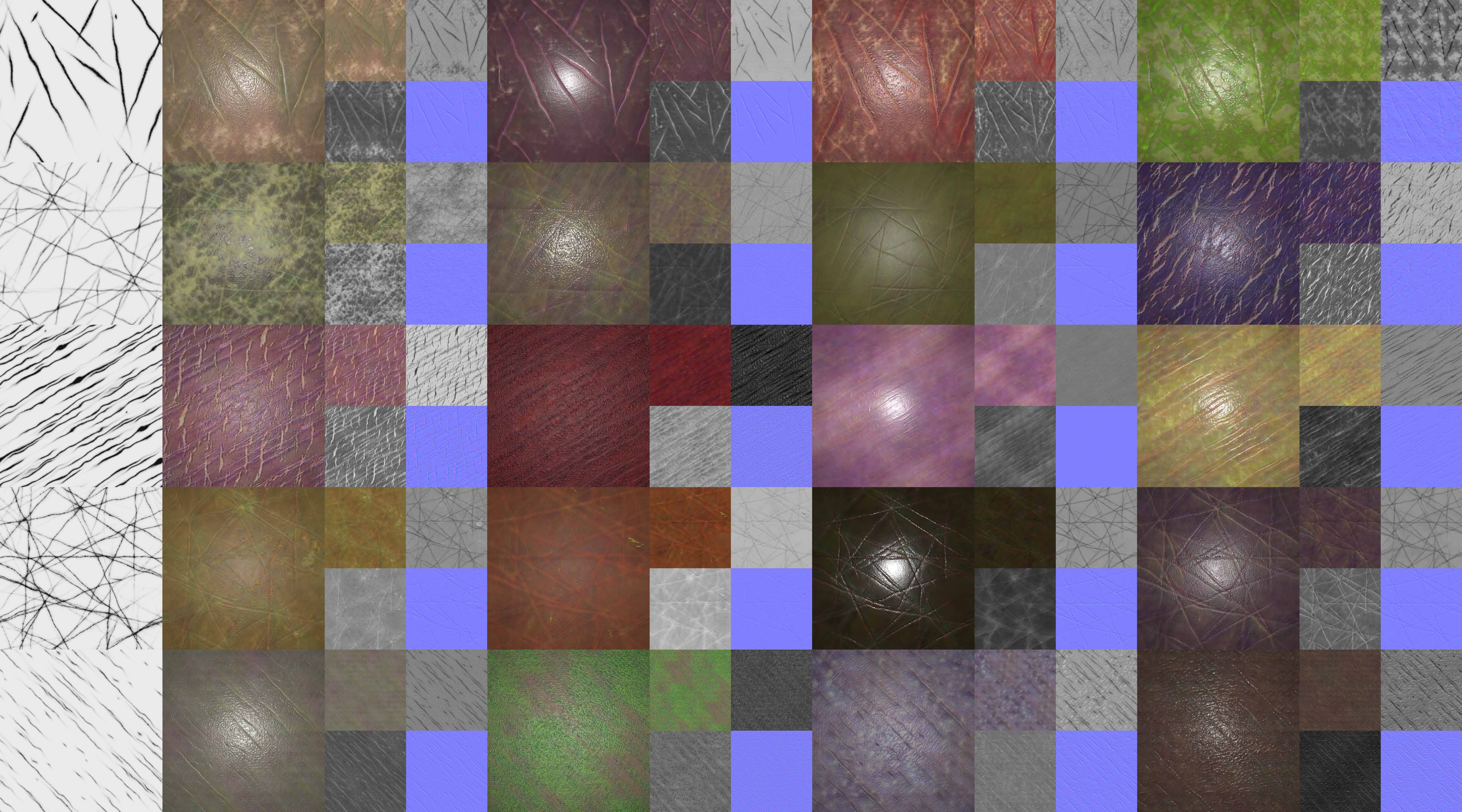}    
\caption{Additional randomly sampled conditional results with the generated texture maps of tile and leather material class.}
\label{fig:cond_supple2}
\end{figure*}

\begin{figure*}[tb]
    \centering
    \includegraphics[width=1\linewidth]{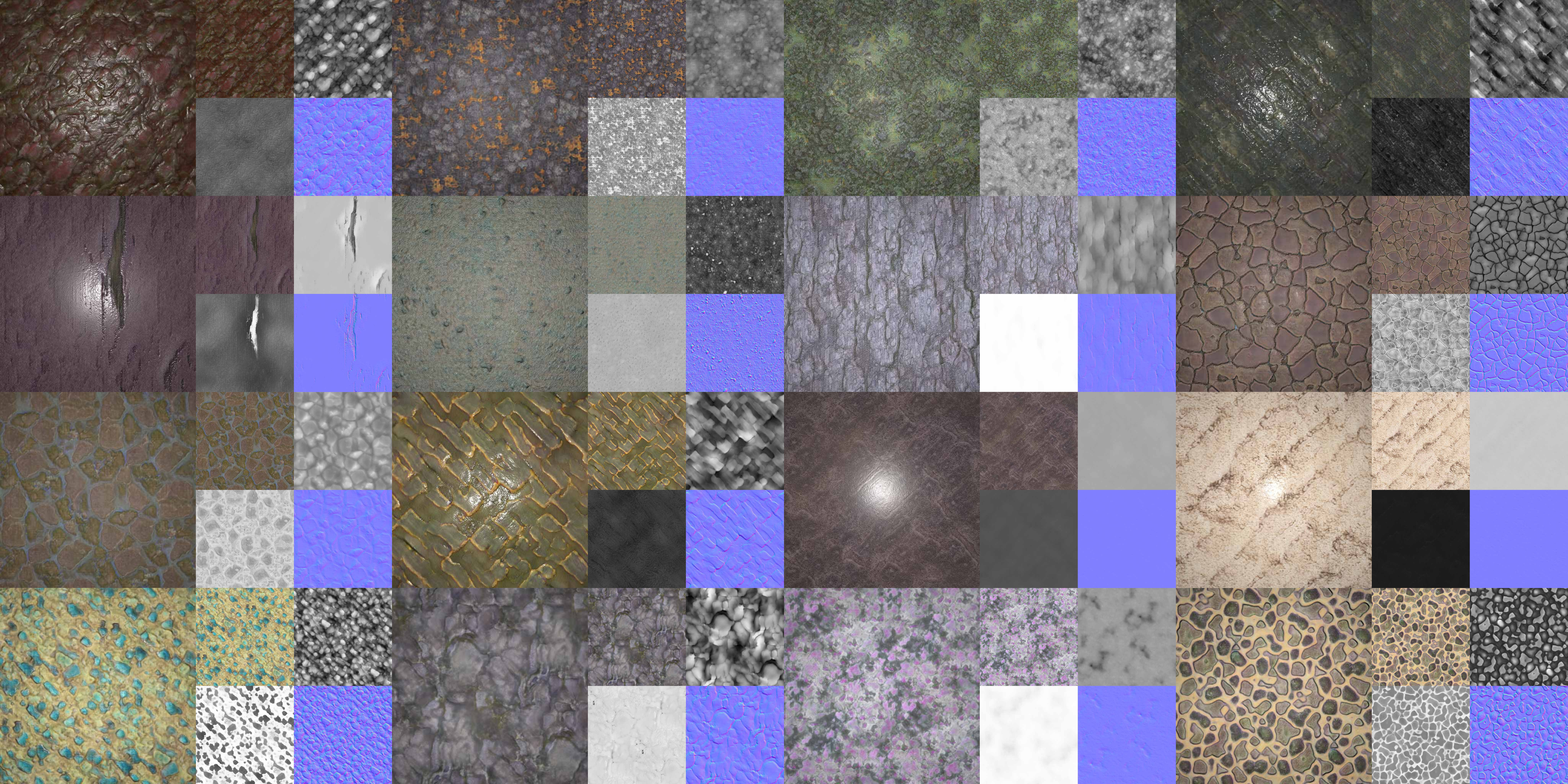}    
    \includegraphics[width=1\linewidth]{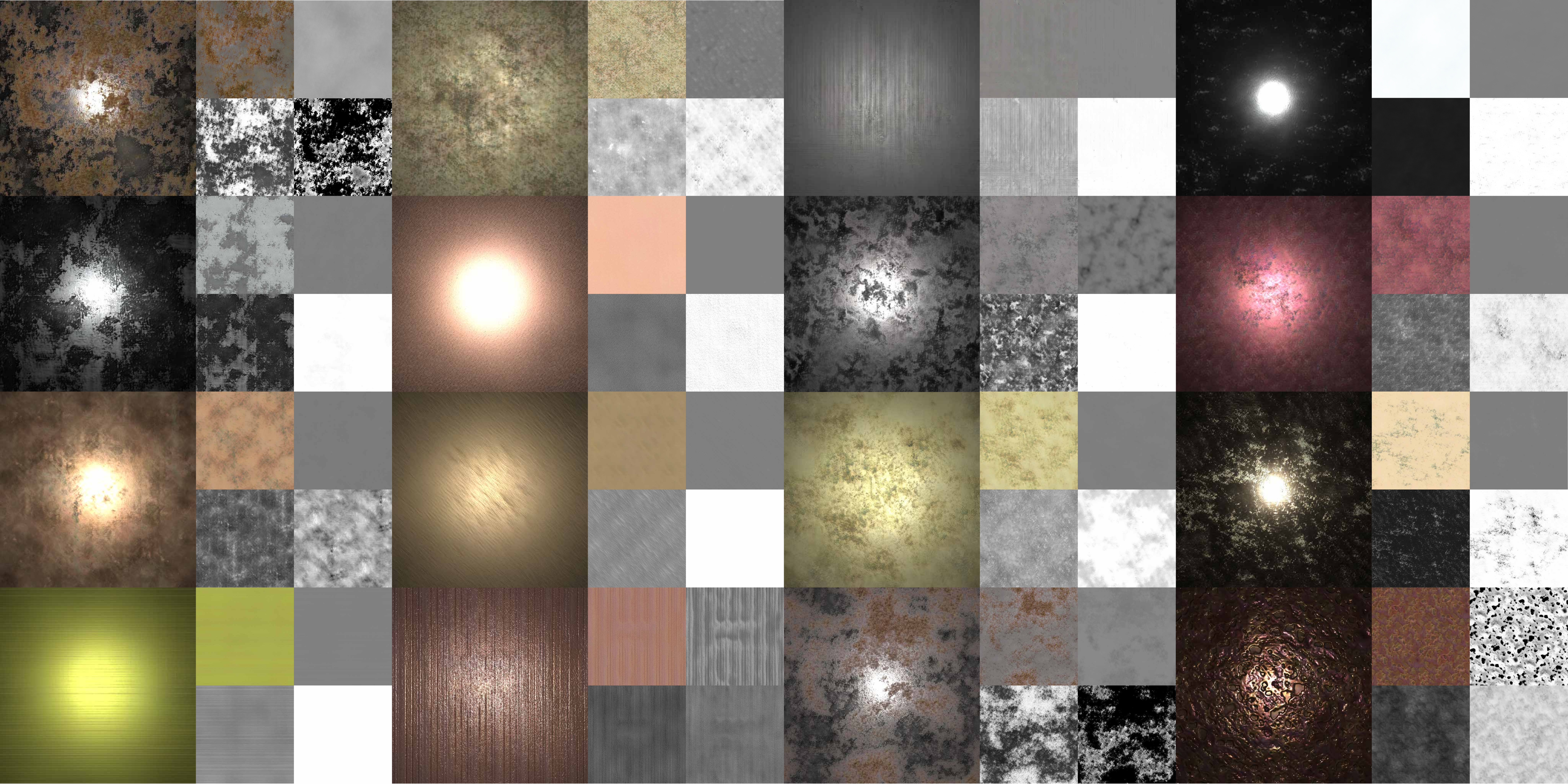}
\caption{Randomly sampled unconditional results with the generated texture maps of stone and metal examples in the Figure 4 of the paper.}
\label{fig:uncond_supple1}
\end{figure*}

\begin{figure*}[tb]
    \centering
    \includegraphics[width=1\linewidth]{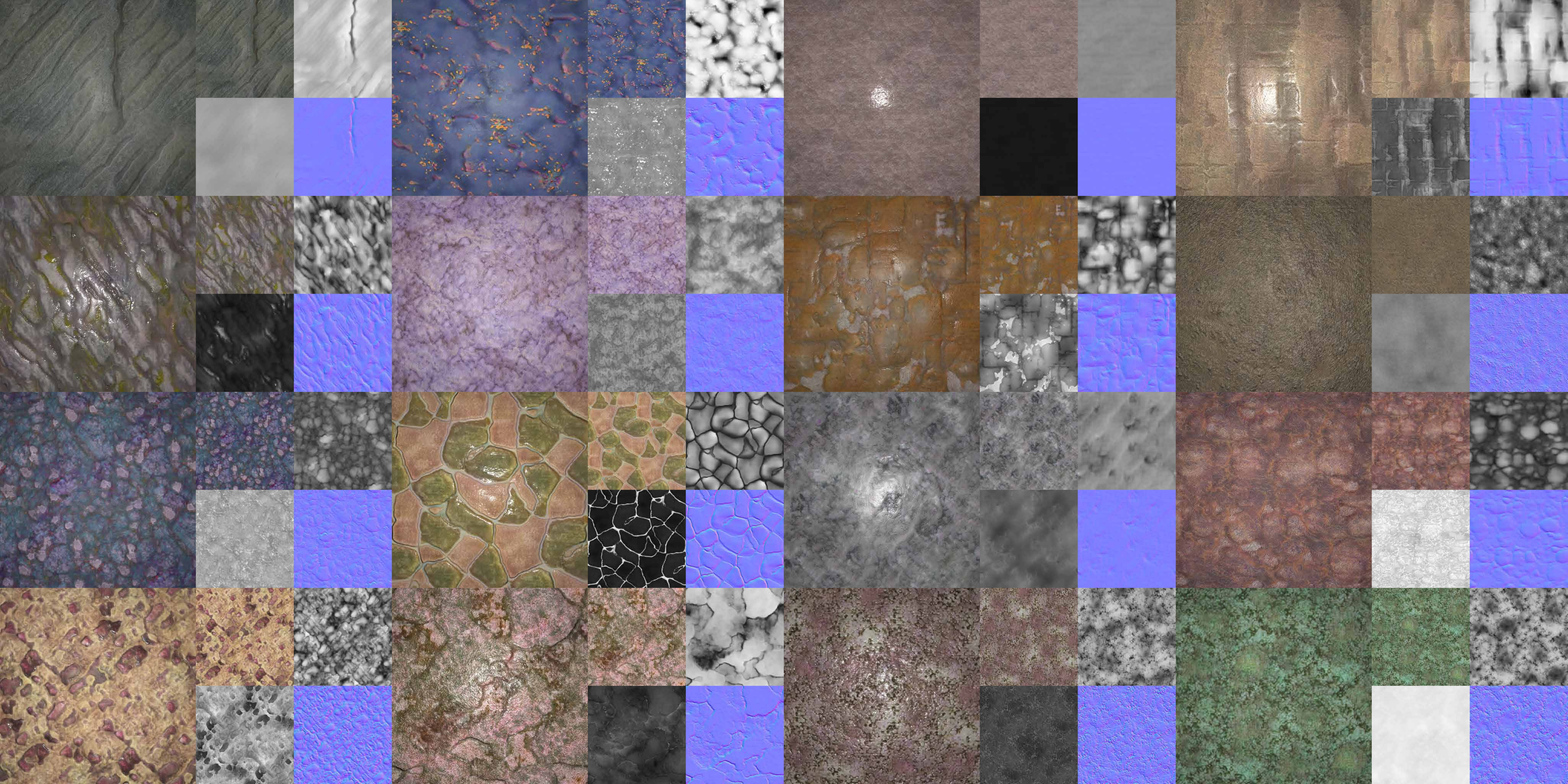}
    \includegraphics[width=1\linewidth]{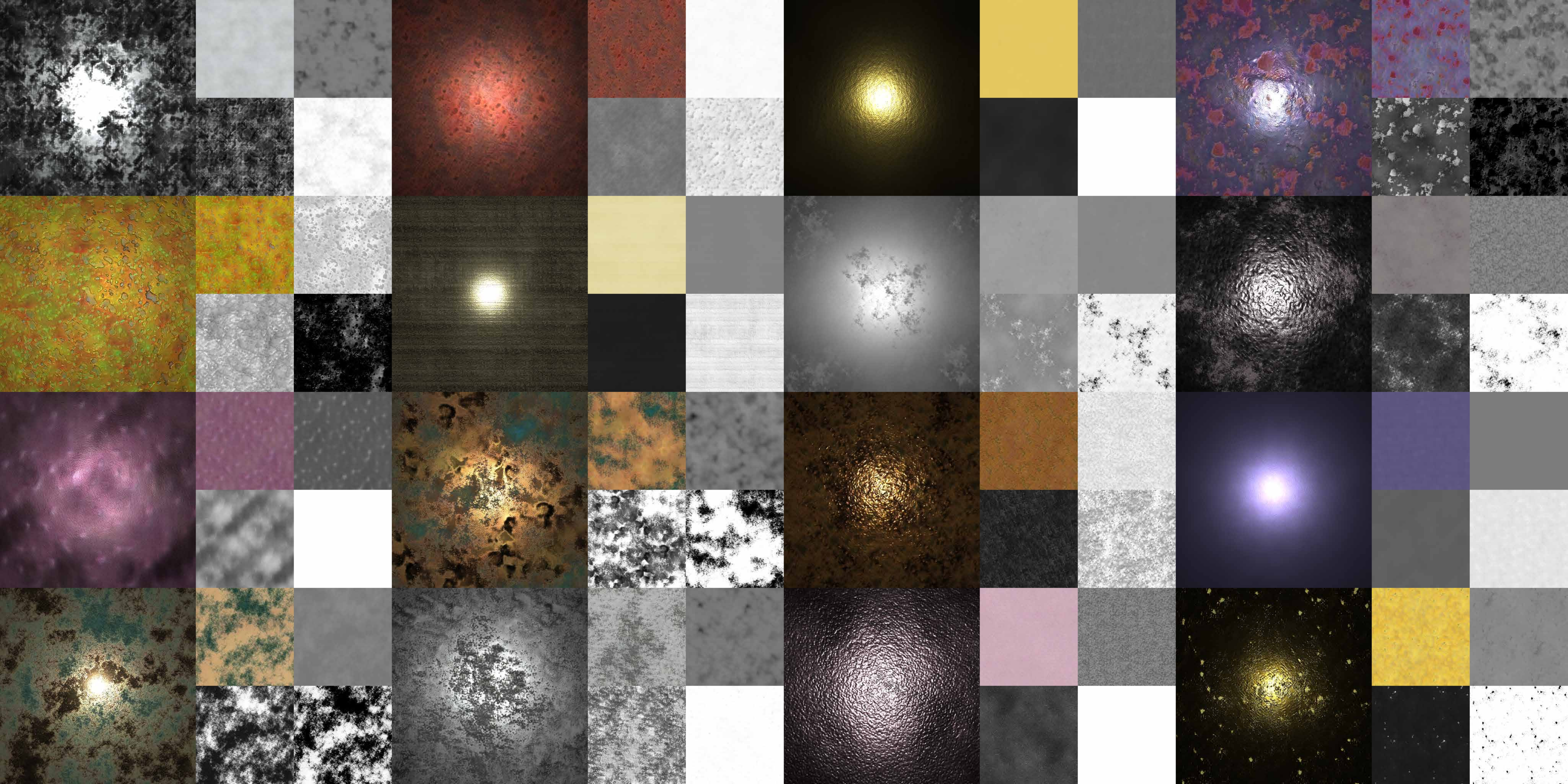}    
\caption{Additional randomly sampled unconditional results with the generated texture maps of stone and metal material class.}
\label{fig:uncond_supple2}
\end{figure*}

\begin{figure*}[tb]
    \centering
    \includegraphics[width=1\linewidth]{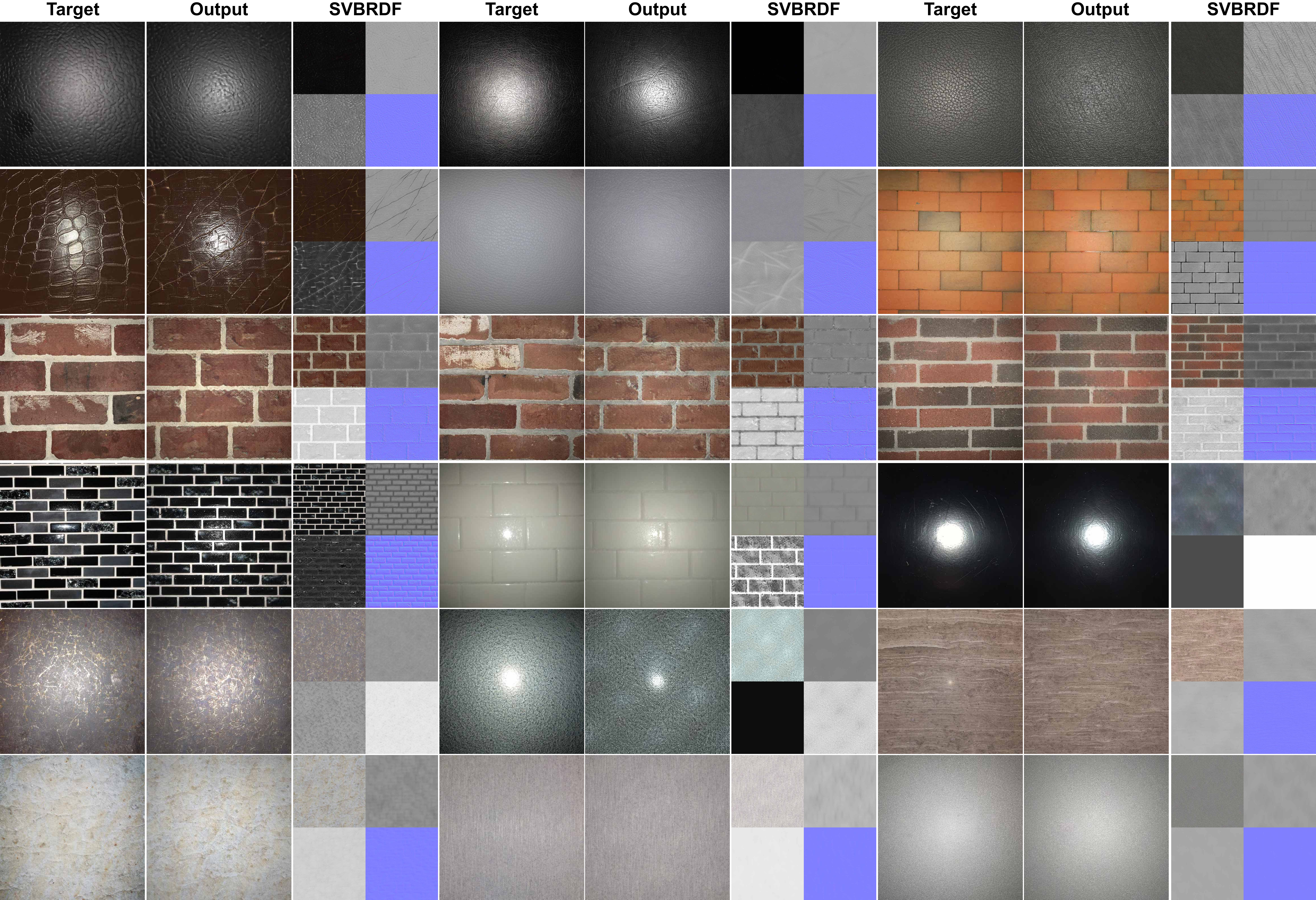}
\caption{Additional inverse rendering results with the generated texture maps.}
\label{fig:inverse_supple}
\end{figure*}

\paragraph{Additional results for generation and inverse rendering.} In Figure~\ref{fig:cond_supple1} we show individual SVBRDF maps for the conditionally generated results shown in Figure 3 of the main paper. More conditionally generated results are shown in Figure~\ref{fig:cond_supple2}. The SVBRDF maps for the unconditional results in Figure 4 of the main paper are shown in~\ref{fig:uncond_supple1}, and more unconditional results in Figure~\ref{fig:uncond_supple2}.
% and~\ref{fig:uncond_supple2}, we show additional results generated without c
% the generated texture maps of randomly sampled materials of Figure 3 and Figure 4 in the paper. In Figure \ref{fig:cond_supple2} and Figure \ref{fig:uncond_supple2}, we show additional randomly sampled materials together with the corresponding texture maps. In Figure 
Figure~\ref{fig:inverse_supple} shows additional inverse rendering results.
%with the corresponding texture maps.

\begin{figure*}[tb]
    \centering
    \includegraphics[width=1\linewidth]{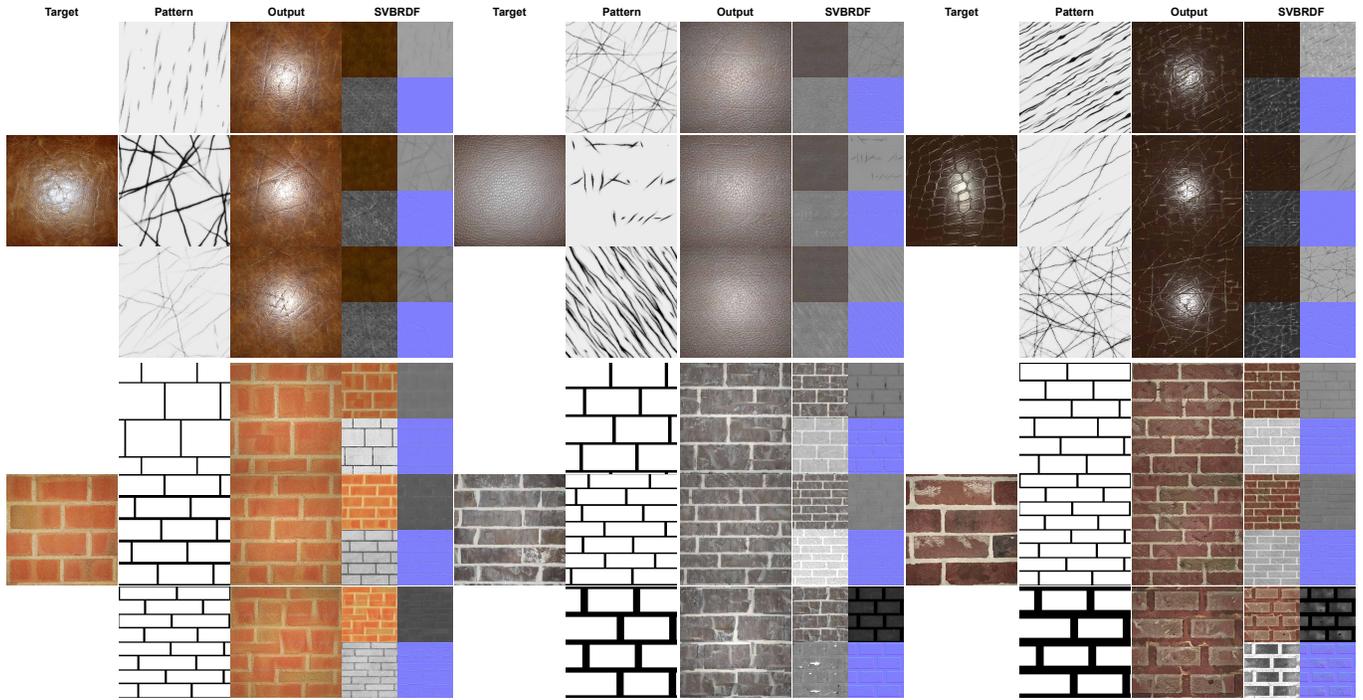}
\caption{This figure shows the effect of conditional input patterns on the generated feature maps given certain target image. For leather examples, our results can capture the style of target images and maintain "wrinkles" introduced from diverse input patterns; for tile examples, given input patterns that are not significantly wrongly aligned with target images, our model can still generate realistic results. }
\label{fig:pat_ablation}
\end{figure*}

\paragraph{Effect of input patterns on inverse rendering} We demonstrate the effect of conditional input patterns on the results of inverse rendering.  In the Figure \ref{fig:pat_ablation}, the top three rows are leather examples and bottom three rows are tile examples. For each example, we match the same target image with 
%show results with
three different input patterns.
%followed by the corresponding renderings and feature maps. 

We can see that in the leather examples, all input patterns produce plausible reconstructions that both capture the style of target image, and also preserving the "wrinkles" specified in the input patterns. In the right-most leather example, the provided wrinkle patterns are significantly different from the wrinkles in the target image, but our result still capture the style of the target while using the wrinkles provided by the input patterns.

The bottom three rows show that for tile examples, the input patterns do not need to be exactly aligned with the target images (Different examples of Figure 5 also demonstrate this property). And between input patterns and target images there exists some "tolerance", including tile number, tile size, the thickness of grout and the offset of tiles. However, if the feature sizes of the input patterns are significantly different from the target images, our model will try to mix style and patterns and generate unrealistic bricks (see the 3rd patterns for each tile example). This is because the Gram matrix loss is shift-invariant but not scale-invariant, and significantly mismatched patterns will force a change in the style computed by the Gram matrix.
%hurt the style coming from target images. 

\begin{figure*}[tb]
    \centering
    \includegraphics[width=1\linewidth]{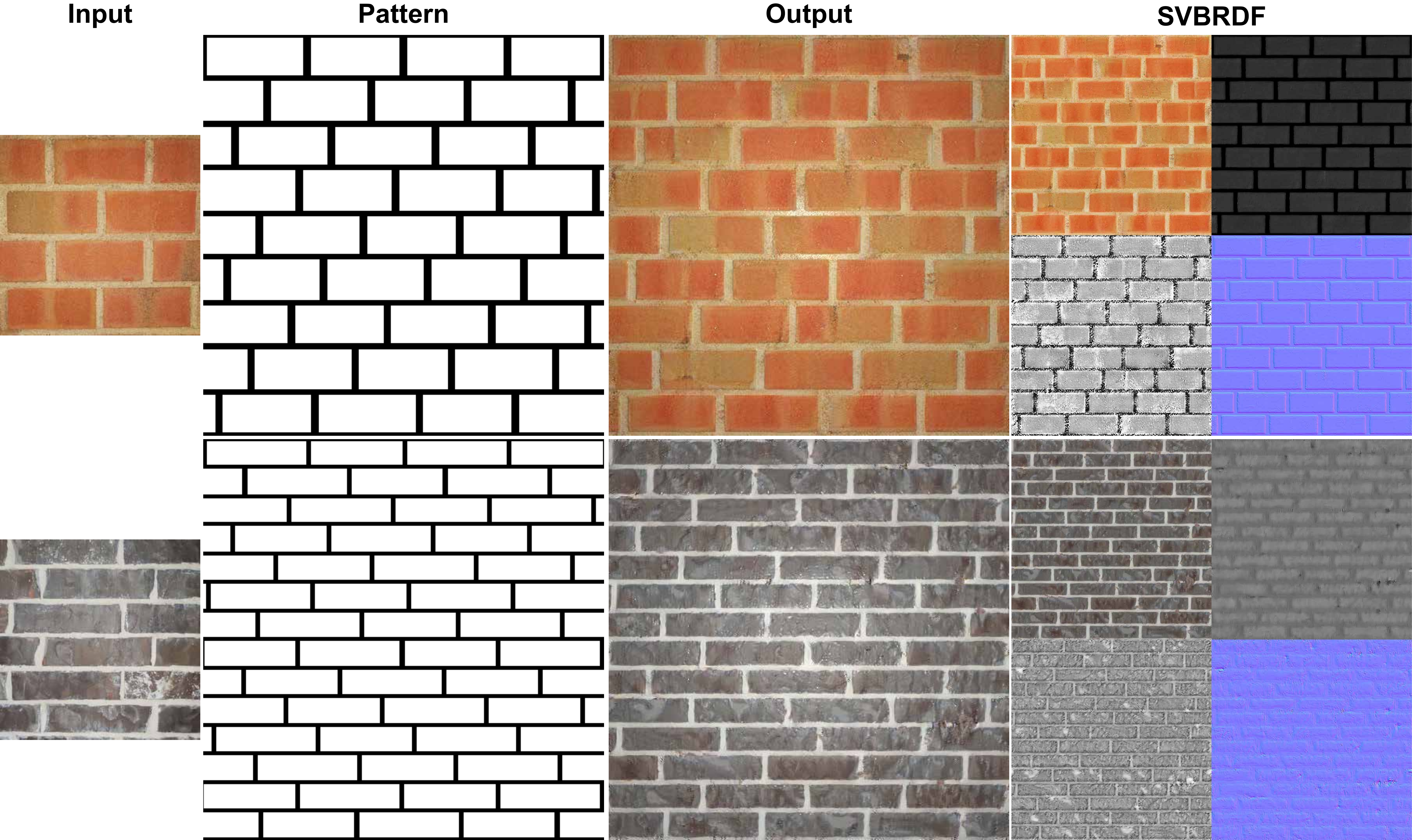}
\caption{The textures resulting from our optimization can be larger than the target image. Here we start from $256 \times 256$ target images, and define a pattern with a feature size matching the target in pixels, but covering a $512 \times 512$ output domain. Our method is able to produce tileable materials of extended size, which is not possible with previous pixel-based methods.}
\label{fig:pat_extension}
\end{figure*}

\paragraph{Results larger than inputs.} Furthermore, we demonstrate that our generated materials can have higher resolution than the target image. In Figure \ref{fig:pat_extension}, we use a $256 \times 256$ target photograph and provide a pattern for a $512 \times 512$ output domain, with a feature size matching the target image in pixel units. Our method results in tileable materials of extended size, which is not possible with previous pixel-based capture methods.

\begin{figure*}[t]
    \centering
    \includegraphics[width=1\linewidth]{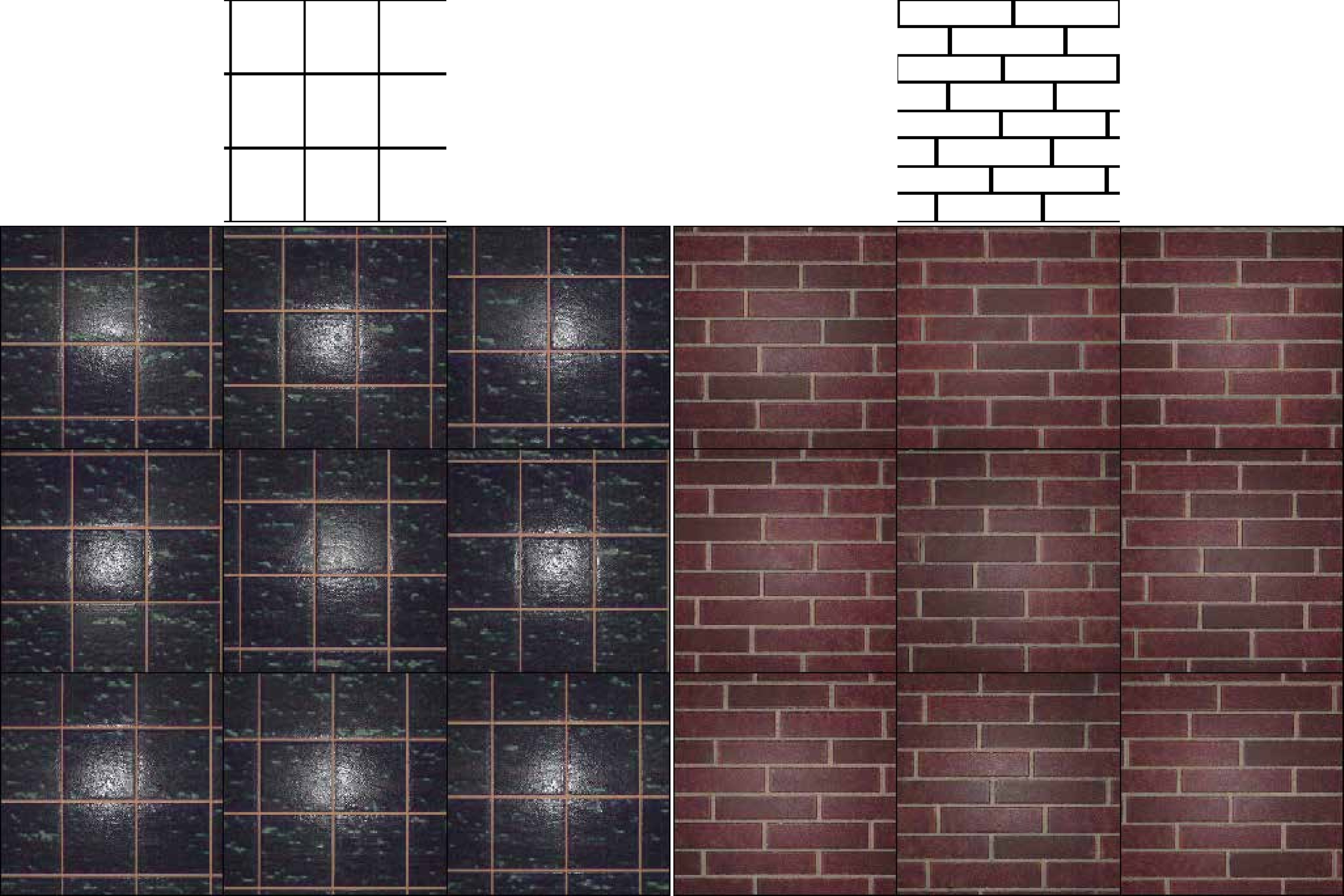}
\caption{A fixed input pattern (shown on the top) is translated by a number of pixels, while keeping a fixed latent code. The resulting material follows the shifted pattern and keeps a similar style, showing that much of the style of the result is derived from the latent vector, not the pattern encoding. This shows that our model provides a certain amount of disentanglement between condition and style, which is sufficient for our forward and inverse tasks.}
\label{fig:fix_style}
\end{figure*}

\paragraph{Translated condition with fixed style.} In Figure \ref{fig:fix_style}, we translate the input pattern, while holding the latent code fixed. The resulting materials follow the shifted pattern but do not change their style, which further demonstrates that the style of the result is mainly derived from the latent vector rather than the pattern encoding. This property is desirable for inverse optimization, where the pattern is fixed and we would like to optimize for the material style. %\KS{can we change the pattern in other ways, otherwise this is not super interesting because we could always just translate the output results directly instead of going via the GAN.}

% \paragraph{Tileability visualization.} The tileability of our inverse rendering results is shown in Figure \ref{fig:Comparision_supple}. The textures resulting from our optimization have been tiled $2 \times 2$ times and rendered, demonstrating seamless tileability (periodicity) of our resulting material maps. We also compare our tiled results with MaterialGAN optimized with same loss as our paper. As is shown in the figure, MaterialGAN optimized with gram matrix loss cannot generate tileable results. 

%Note that real-world target images are typically not tileable, and this is not an issue for our approach.

% \paragraph{Latent space interpolation.} In Figure \ref{fig:interpolate}, we show latent space interpolation between two random sampled latent space for tiles and leather, while keeping a fixed condition pattern. \xz{maybe remove this part?}

\begin{figure*}[t]
    \centering
    \includegraphics[width=1\linewidth]{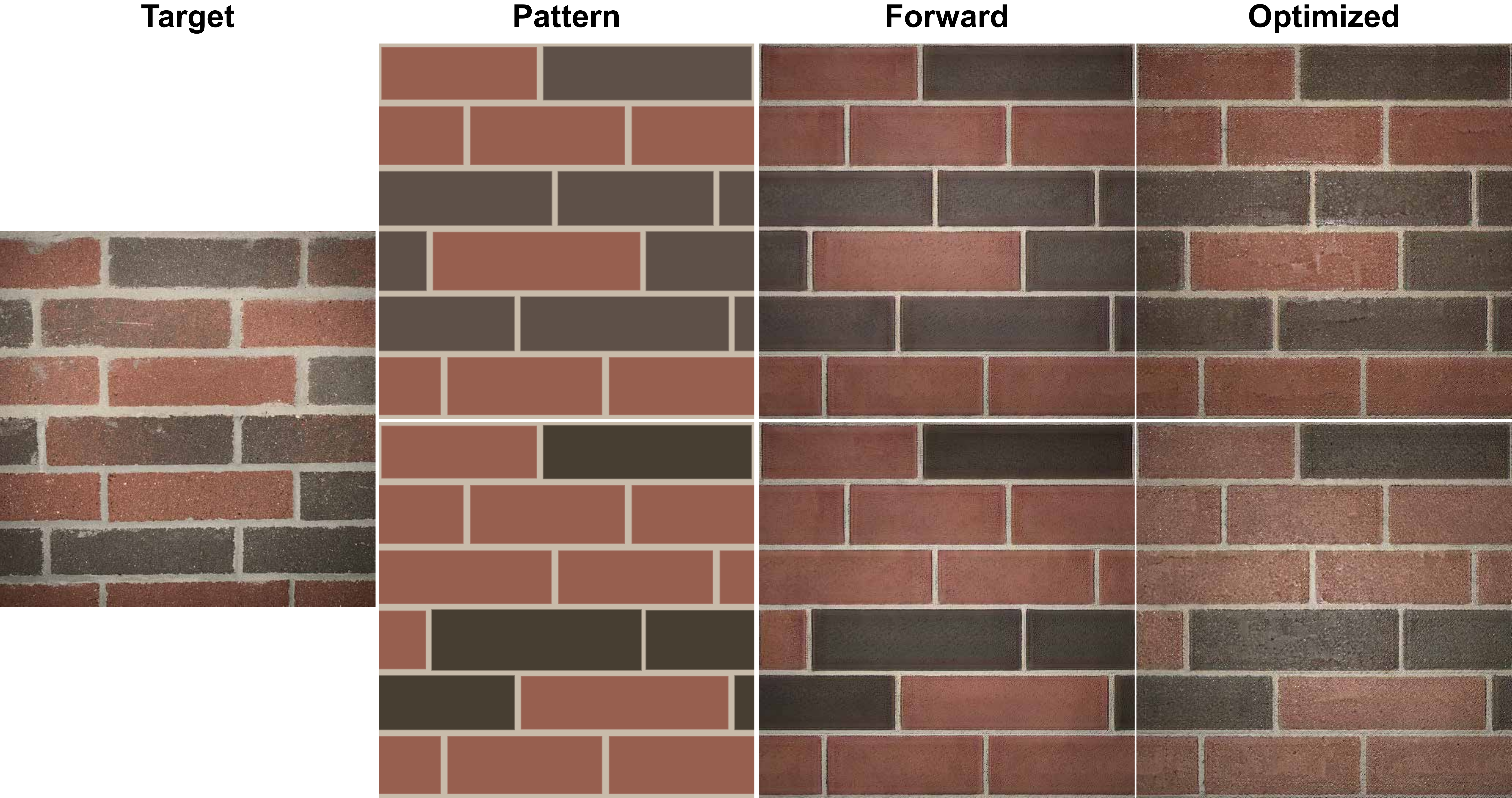}
\caption{Here we demonstrate the color version of our conditional generator. We provide a pattern with roughly matching colors to the generator (forward) and optimizer (Optimized). Even though the colors provide significant information which doesn't align well with the input picture, the optimization is able to further match the style of the target photograph.}
\label{fig:color_pat}
\end{figure*}

\paragraph{Conditioning on colored patterns.} In addition to conditioning on binary patterns, we can also train our generator to allow conditioning on colored patterns to control the diffuse albedo in the generated materials ---other properties of other materials classes could be conditioned this way. 
%Apart from conditioning by binary patterns, our model can also be trained conditioned on color patterns, where the color in different tiles can vary, controlling the average diffuse albedo in the final textures.
In Figure \ref{fig:color_pat}, we provide a pattern with colors roughly matching the target photograph, but at different image locations. The generator produces a random style that matches the given colors. Optimization to match the target image further reproduces the detailed style of the target photograph, even though the colors in the pattern already provide significant prior information.

% \paragraph{Tileable architecture ablation.} We show in Figure~\ref{fig:ablation_tile} an analysis of a result with and without the modifications we perform on the loss and network architecture. Since we design our approach to be translation-invariant and prevent it from recognizing the position of any image patch relative to the border, we do not suffer from tiling artifacts. The simpler alternative would be to not modify the architecture to be tileable, but still train it on tileable examples; this performs worse than our architecture and still shows clear tiling artifacts.
% %\paragraph{Effect of tileable generator}  See Fig \ref{fig:ablation_tile} 

\newpage

% \begin{figure*}[tb]
%     \centering
%     \includegraphics[width=1\linewidth]{img/sampling_cond.pdf}
% \caption{We feed the conditional pattern on the left to \cmgan, along with four different random latent vectors $\bz_1,\dots,\bz_4$, to produce corresponding material instances. The results have varied appearance but the layout of the condition. The corresponding texture maps are shown in supplementary materials. }%Figure S\ref{fig:cond_supple1}. }
% \label{fig:rand_sample_cond}
% \end{figure*}

% \begin{figure}[t]
%     \centering
%     \includegraphics[width=1\linewidth]{img/interpolate.pdf}
% \caption{The results of style interpolation between two random sampled latent space for tiles and leather.}
% \label{fig:interpolate}
% \end{figure}

% \begin{figure}[tb]
%     \centering
%     \includegraphics[width=1\linewidth]{img/ablation_tile.pdf}
% \caption{Our tileable architecture clearly outperforms the alternative of training on tileable data but without the architecture modifications.}
% \label{fig:ablation_tile}
% \end{figure}

% \bibliographystyle{ACM-Reference-Format}
% \bibliography{references}

% \end{document}